\documentclass{jfm}
\usepackage{graphicx}
\usepackage{epstopdf, epsfig}
\graphicspath{{texfig/}}
\usepackage{subfigure}
\usepackage{mathtools}
\usepackage{amsmath}
\usepackage{amssymb}
\usepackage{float}
\usepackage{verbatim}
\usepackage[titletoc]{appendix}
\usepackage{booktabs} 
\usepackage{color}
\usepackage{apacite}
\usepackage{natbib}

\newcommand{\mybar}[1]{\makebox[0pt]{$\phantom{#1}\overline{\phantom{#1}}$}#1}
\renewcommand{\d}{\textnormal{d}}
\newcommand{\D}{\textnormal{D}}
\newcommand{\RNC}[1]{\MakeUppercase{\romannumeral #1}}
\makeatletter
\@addtoreset{equation}{section}
\makeatother
\renewcommand{\theequation}{\arabic{section}.\arabic{equation}}

\title{The internal structure of forced fountains}

\author{Jingzi Huang
  \corresp{\email{jingzi.huang17@imperial.ac.uk}},
  Henry C.\ Burridge
 \and Maarten van Reeuwijk}

\shortauthor{Huang, Burridge and Van Reeuwijk}

\affiliation{
Department of Civil and Environmental Engineering, \\ Imperial College London, London SW7 2AZ, UK}
\begin{document}

\maketitle

\begin{abstract}
We study the mixing processes inside a forced fountain using data from direct numerical simulation. The outer boundary of the fountain with the ambient is a turbulent/non-turbulent interface. Inside the fountain, two internal boundaries, both turbulent/turbulent interfaces, are identified: 1) the classical boundary between upflow and downflow which is composed of the loci of points of zero mean vertical velocity; and 2) the streamline that separates the mean flow emitted by the source from the entrained fluid from the ambient (the separatrix). We show that entrainment due to turbulent fluxes across the internal boundary is at least as important as that by the mean flow. However, entrainment by the turbulence behaves substantively differently from that by the mean flow and cannot be modelled using the same assumptions. This presents a challenge for existing models of turbulent fountains and other environmental flows that evolve inside turbulent environments.
\end{abstract}


\section{Introduction}
\label{Sec:F1_introduction}

Understanding the process of turbulent mixing within fluid flows is a ubiquitous challenge in the modelling of environmental processes, and in our exploitation of similar physical processes to meet societal needs and demands \citep{Fernando1991}. Frequently, the understanding of the process of turbulent mixing between distinct bodies of fluid is approached by seeking to investigate the exchanges of the physical properties between these fluid bodies. In some circumstances, the region where these exchanges occur is robustly defined by the constraints imposed by the presence of solid boundaries; with the fluid behaviour in the proximity of that region being significantly altered by these boundaries. Examples range in scale, at least, from the flow through the windows of our buildings \citep{Gladstone_woods_2001} to the regions that occur when continental landmasses and the ocean floor conspire to create so-called `straits', which affects the dynamics of our global oceans \citep{Finnigan1999,Finnigan2000}. However, an even wider range of exchanges between distinct bodies of fluid can be driven by turbulence in the absence of any solid boundary. The study of such exchanges is long-standing and ongoing \citep[see][for a broad review]{Dimotakis2005}. It is this type of exchange, specifically that within a canonical turbulent forced fountain, which forms the focus of this study.

The term `turbulent fountain' describes the flow in which the buoyancy force opposes the initial momentum flux of an ejection from a localised source; for example, a negatively buoyant high Reynolds number jet continuously injected upwards into a uniform quiescent environment will produce a turbulent fountain \citep{Hunt2015}. In this frame of reference, quasi-steady fountains consist of an inner core, herein the `upflow', shrouded by a returning counterflow, herein the `downflow' \citep[see, for example][]{Turner_1966,McDogall1981}. Amongst other applications \citep[see][]{Hunt2015}, turbulent fountains play important roles in determining the environment within our buildings \citep[e.g.][]{Linden2005}, water recirculation and purification \citep[e.g.][]{bloomfield_kerr_2000}, and the forced heating or cooling of enclosed spaces \citep{baines1990}. The dynamics of high-Reynolds number high-P\'eclet number miscible fountains can be described by consideration of a single dimensionless parameter \citep{burridge_hunt_2014}: the source Froude number 
    \begin{equation}
            Fr_0 = \frac{{M_0}^{5/4}}{Q_0 {|F_0|}^{1/2}} = \frac{w_0}{\sqrt{|b_0| r_0}},  
    \end{equation}
where $Q_0, M_0, F_0$ are the integral source fluxes, per unit $\pi$, of volume, momentum, and buoyancy, respectively; $b_0 = F_0/Q_0$ is the source buoyancy scale, $r_0 = Q_0/M_0^{1/2}$ is the source radial scale, and $w_0 = M_0/Q_0$ is the characteristic vertical velocity at the source. Note that $r_0$ is the physical radius of the source only if one takes the velocity profile at the source to be uniform (i.e. top-hat). These fountains: a) offer a rich variety of turbulent exchanges across non-solid boundaries, b) are canonical flows and as such can inform our fundamental understanding of the physics of these processes, and c) have been relatively unexplored, compared with other canonical flows, like plumes or jets.

Exchanges driven by turbulence, in the absence of a solid boundary, have traditionally been studied from perspectives driven by two distinct motivations. Firstly, that of understanding, and often modelling, the fundamental physics for these processes, e.g., the modelling of gravitational convection of turbulent plumes and jets \citep[e.g.][]{Morton1956}; or secondly, from a need to account for the effects of these physical processes within societal applications, e.g., the modelling of atmospheric dynamics for weather prediction \citep[for example][]{Neggers2002}. These different motivations have led to subtly different meanings of the same key terms within different fields of scientific literature. As such, we devote some effort to clarifying, first the history, and then our usage of these terms, in particular `entrainment'. 

When driven by the desire to develop widely applicable models for the effects of turbulence, the term `entrainment' has classically referred exclusively to the drawing in of mass (volume) by a region of turbulent fluid from an ambient background \citep{Morton1956}; implicitly this occurs across what is now termed a `turbulent/non-turbulent interface (TNTI)'. This physical process is relevant to a broad class of canonical flows, e.g., jets \citep{Abramovich1984}, plumes \citep{PB1955,Morton1956}, temporal jets \citep{Van_holzner_2014}, clouds \citep{deRooy2013}, growing regions of turbulence driven by an oscillating grid \citep{Thompson_turner_1975}, and between fountain flows and the environment at their envelope \citep{burridge_hunt_2016}; for a thorough review see \citet{daSilva2014}. By definition, a quantity that is exchanged from the background into the region of interest is referred to as being entrained, or `entrainment'; if it leaves the region of interest it is often referred to as being detrained or `detrainment' --- we adhere to this usage of these terms.

In the context of these scientific endeavours, the term `turbulent entrainment' has historically been used to emphasise the physical mechanism by which this process occurs. However, this `entrainment' is classically associated with a mean inflow velocity across the TNTI which is related to a characteristic velocity of the flow within the region of interest, in this case, the only region of turbulent fluid, via an entrainment coefficient \citep{turner_1986}. This inflow might not only be associated with fluxes of mass (volume), but also with other quantities such as momentum and buoyancy \citep{Talluru2022}. This can be insignificant, as in the canonical cases of plumes or jets when the background is a quiescent uniform environment, but becomes significant, for example, whenever the background is stratified or in motion. These considerations are pertinent to the modelling of the momentum and buoyancy exchanges between the upflow and downflow within fountains \citep[][hereafter referred to as BK00 and SH14, respectively]{bloomfield_kerr_2000, shrinivas_hunt_2014}.

More recently, some studies have focused on cases where the background fluid is also turbulent (rotational) so that the exchanges occur across a turbulent/turbulent interface (TTI) \citep{kankanwadi_buxton_2020, Maarten_2021}. In cases that feature a TTI, the use of the term `turbulent entrainment' referring to the transport by the mean properties of a turbulent flow region is no longer helpful, since the flow is turbulent on both sides of the interface (TTI).

An unambiguous terminology is thus needed to describe the different exchanges when dealing with TTIs. We will refer to the exchanges into our region of interest (which must be defined explicitly in each case) due to the mean flow properties as `mean entrainment' and the exchanges due to turbulence as `turbulent entrainment'. However, to avoid a conflict with the classical usage of the terms, we shall not use the term `turbulent entrainment' in isolation, instead we always explicitly specify the quantity that is being exchanged, e.g.\ `mean volume entrainment' and `turbulent momentum entrainment', etc.

The aforementioned entrainment terms can be defined rigorously using the example of a slender axisymmetric steady-state jet/plume/fountain developing in the direction of the coordinate $z$, for which the integral continuity equation and transport equation of a generic scalar, or vector component, $X$ in absence of sources are given by \citep[see][]{Maarten_2021}
\begin{align}
  \label{eq:F1_contintavGS}
 \frac{\d}{\d z} \int_{\overline{\Omega}} \overline{w}   ~ \d A 
 &= -\oint_{\partial \overline{\Omega}}  \mybar{V}_{g} ~ \d \ell, \\
  \label{eq:F1_scalavGS}
\frac{\d}{\d z} \int_{\overline{\Omega}} \left( \overline{w} \overline{X} + \overline{w'X'} \right) ~ \d A 
&= - \oint_{\partial \overline{\Omega}}  \left( \mybar{V}_{g} ~\overline{X} + \overline{V_{g}' X'} \right)  ~ \d \ell.
\end{align}
Here, the velocity component in the $z$-direction is denoted $w$, and integration is performed over a domain $\overline{\Omega}$, orthogonal to the $z$-direction, that encompasses the flow within the region of interest; $\partial \overline{\Omega}$ denotes the boundary of $\overline{\Omega}$. The overbar indicates the domain is chosen based on a Reynolds-averaged quantity: in the case of jets/plumes, this Reynolds-averaged quantity is typically a small fraction $\epsilon$ of the centreline velocity $\overline{w}_c$ \citep{Van2016}, i.e. $\overline{\Omega}$ is the domain for which $\overline{w}>\epsilon \overline{w}_c$. Since the flow is statistically axisymmetric and time-independent, the interface can equally be described using a single-valued function $r=h_t(z)$.

The instantaneous entrainment velocity across the interface is denoted $V_g$, which is the difference between the local fluid velocity and the boundary velocity. Following \citet{Maarten_2021}, we write this as
\begin{equation} \label{eq:F1_V_g}
    V_g = u - \frac{\D h_t}{\D t} = u - w \frac{\d h_t}{\d z} \, , 
\end{equation}
where $\D h_t/\D t$ is the material derivative representing the boundary velocity, $u$ and $w$ are the \emph{instantaneous} radial and streamwise velocity components of fluid on the interface $\partial \overline \Omega$, respectively. For the case under consideration, $h_t$ is time-independent due to consideration of a boundary defined by a Reynolds-averaged quantity, in which case this term then simplifies as per the right-hand side (RHS) of (\ref{eq:F1_V_g}). 
Reynolds-averaging the expression above results in $\overline{V}_g = \overline{u} - \overline{w} \frac{\d h_t}{\d z}$.
Note that since we choose the interface based on a threshold of velocity, the last term is negligibly small compared to the first, and we recover the classical entrainment velocity $\mybar{V}_g = \overline{u}$ for jets/plumes.

The RHS of (\ref{eq:F1_contintavGS}) represents the \emph{mean volume entrainment}, as the transport is due to the mean quantity $\mybar{V}_g$, in this case $\mybar{V}_g = \mybar{u}$. The first term on the RHS of (\ref{eq:F1_scalavGS}) represents the \emph{mean scalar entrainment} term $\mybar{V}_g \overline{X}$ and the second term of the RHS is the \emph{turbulent scalar entrainment} term  $\overline{V_{g}' X'}$ (by definition $X$ can also be a vector component, e.g. $w$; we omit the `vector' in the name here for simplicity). Note that volume entrainment can only be caused by the mean flow, but \emph{all} other transported properties (momentum, buoyancy, etc.) can have both a mean and a turbulent component.

\begin{figure}
	\centering{\includegraphics[width=7.5cm]{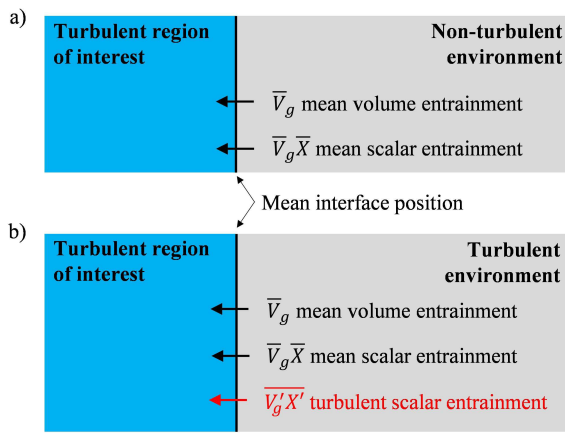}}
	\caption{Schematic illustrating the entrainment of volume and scalars across interfaces separating the flow within the chosen region of interest, which is by definition turbulent, and the environment. Two different cases emphasise our chosen terminology: a) a non-turbulent environment, and b) a turbulent environment. Transport or exchanges across the interface, here all termed `entrainment', are illustrated by arrows; for each entrainment, the associated transport term and its naming convention are highlighted. Black arrows represent the entrainment that is associated with mean flow, while the red arrow represents that associated with turbulence.}
	\label{fig:F1_terminology}
\end{figure}

The terms for entrainment are illustrated schematically in figure~\ref{fig:F1_terminology}. The figure illustrates the entrainment from an ambient of two cases: a) the ambient is non-turbulent so that the interface is a TNTI, for which only mean volume entrainment and mean scalar entrainment can occur at the interface (e.g. the interface between a turbulent fountain and the quiescent environment); b) the ambient is turbulent so that the interface is a TTI (e.g. the interface between the upflow and downflow inside a turbulent fountain), for which, besides mean volume entrainment and mean scalar entrainment, turbulent scalar entrainment can occur at the interface. Instantaneous exchanges at the TTI can and do occur in both directions \citep[as allowed in the BK00 model of a fountain or measured at cloud edge][]{deRooy2013}, but the Reynolds averaged quantities discussed here are the net value of these bi-directional exchanges. 

In some circumstances, the fluid on one particular side of the TTI has a larger turbulence intensity and so the choice of the `region of interest' (or `flow region') and `background' is obvious. However, for more complex fluid flows this is not so straightforward. For example, in a fountain one can distinguish between an upflow and a downflow, which are modelled separately (BK00). Here, in the upflow and in the downflow, turbulence levels can be of comparable turbulence intensity, with standard metrics switching from being typically larger in the upflow near the source, to the downflow further from the source. Another example is a cumulus cloud, where it is also difficult to distinguish on which side of the interface between the cloud core and descending shell the turbulence intensity is largest \citep{deRooy2013,Heus2008,Vishnu2020}. Thus in order to distinguish `entrainment' and `detrainment', one must always clearly define the `region of interest', as this will determine whether one should use the term entrainment (exchange into the region) or detrainment (exchange out of the region). For completeness, we note that should the exchanges illustrated in figure~\ref{fig:F1_terminology}(b) be evaluated to be negative then, following our convention, this would result in the reversal of the direction of the arrows, and the exchanges would thus be termed `detrainment'.
 
\vspace{0.5\baselineskip} 
The aim of this work is to study the internal structure of a turbulent forced fountain; to do so, we use direct numerical simulation (DNS) to investigate a highly forced fountain. The structure of the paper is as follows: in \S \ref{Sec:F1_simulation_details} we describe the numerical simulation details. In \S \ref{Sec:F1_internal_structure} we describe the instantaneous fountain observation and discuss the internal structure of a Reynolds-averaged fountain, we divide the fountain into three regions of interest with two turbulent/turbulent interfaces. In \S \ref{Sec:F1_equations} we introduce the integral quantities of a fountain and appropriate conservation equations in order to find the flux and entrainment budgets of mass, momentum and buoyancy. In \S \ref{Sec:F1_Integral_quantities} we examine the vertical variation in fluxes within the three flow regions. In \S \ref{Sec:F1_transfer} we discuss entrainment across the interfaces respectively, including that near the fountain top; ending with a comparison of the entrainment coefficients from our DNS to those of other studies. Concluding remarks are made in \S \ref{Sec: F1_conclusion}.

\section{Simulation details} \label{Sec:F1_simulation_details}

\begin{figure}
	\centering{\includegraphics[scale = 0.6]{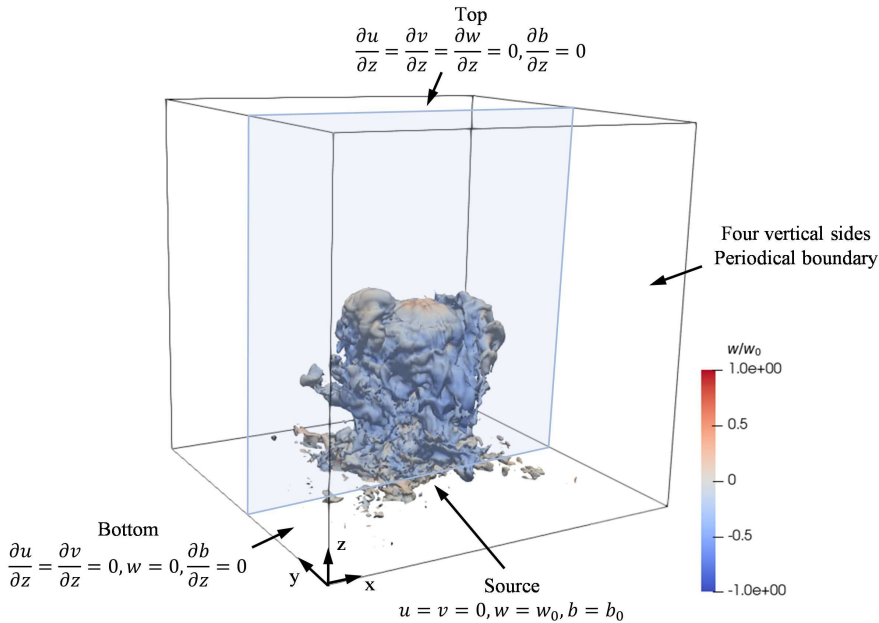}}
	\caption{The whole simulation domain with an illustration of an instantaneous fountain depicted by plotting an isosurface of buoyancy, coloured according to the local vertical velocity. The unit of the length is the number of nodes.}
	\label{fig:F1_simulation_domain}
\end{figure}

We simulate an axisymmetric fountain driven by an isolated source of uniform (top-hat) vertical upwards velocity $w_0$ and opposing (negative) buoyancy $b_0$ where buoyancy is defined as $b = g(\rho_e - \rho)/\rho_a$, in which $\rho$ and $\rho_e$ denote the density of fountain flow and environment, respectively, and $\rho_a$ is a reference density. The domain has size  $L_x \times L_y \times L_z = 160\,r_0 \times 160\,r_0 \times 100\,r_0$ where $r_0$ is the circular source radius.
The coordinate system has its origin in the bottom corner with the $z$-direction upwards (figure~\ref{fig:F1_simulation_domain}), and the source is located in the centre  at  $(L_x/2, L_y/2, 0)$. The source fluxes, per unit $\pi$, of volume $Q_0$, momentum $M_0$, buoyancy $F_0$, and (again per unit $\pi$) the integral buoyancy $B_0$, can be defined as
\begin{equation}
    Q_0 = w_0r_0^2, \quad M_0 = w_0^2r_0^2, \quad F_0 = w_0b_0r_0^2, \quad B_0 = b_0r_0^2.
\end{equation}
The fountain source is such that the source Froude number is $Fr_0 = 21$ which sits well above the threshold ($Fr_0 \gtrsim 4.0$) for the region described as forced fountains \citep{burridge_hunt_2012}. The Reynolds number is $Re_0 =w_0r_0/\nu=1\,667$, sufficient such that our simulations sit within a Reynolds number regime ($Re_0 \gtrsim 1\,000$) for which the experiments of \citet{burridge2015effect} demonstrate that the bulk behaviour of the fountain, in particular the fountain rise height, is broadly independent of $Re_0$, and the flow can be regarded as high Reynolds number, or `turbulent'. 

For these turbulent highly forced fountains universal scalings exist based on consideration of a point source fountain of momentum flux, $M_0$, and opposing buoyancy flux, $F_0$. By consideration of the length scale $L_{F} = M_0^{3/4}/|F_0|^{1/2}= r_0 Fr_0$ \citep[\textit{cf.}][who originally defined the length scale based on physical fluxes, rather than the fluxes per unit $\pi$, i.e., there is a ratio of $\pi^{1/4}$ between the two length scales]{Turner_1966} and time scale $T_{F} = M_0/|F_0| = \sqrt{r_0/|b_0|} \,  Fr_0$ \citep{burridge_hunt_2013}, results for forced fountains are therefore universal when scaled on $Fr_0$. Herein, we use these scales to normalise all quantities, making our findings directly applicable to all turbulent fountain flows with source Froude number $Fr_0 \gtrsim 4$ \citep{Hunt2015}. For example, we normalise all volume fluxes using $Q_F = L_F^3 / T_F = M_0^{5/4}/|F_0|^{1/2}$ \citep[$Q_F \propto Q_0 Fr_0$][]{baines1993} and for all quantities concerning the entrainment of volume, momentum and buoyancy, we use $q_F = Q_F/L_F = M_0^{1/2}$, $m_F = M_0/L_F$, and $f_F = F_0/L_F$, respectively. The size of our cuboidal domain in these units, is $L_x \times L_y \times L_z  = (7.55\, L_F)^2 \times 4.72\, L_F$.

The incompressible Navier-Stokes equations under the Boussinesq approximation are solved numerically using DNS on a uniform Cartesian grid of $N_x \times N_y \times N_z = 1280^3$ cells. The grid size is resolved smaller than two Kolmogorov scales, such as $\Delta x = L_x/N_x < 2\eta_K$. Here the Kolmogorov scale $\eta_K=(\nu^3/\epsilon)^{1/4}$, where $\epsilon$ is the averaged integral dissipation rate of turbulent kinetic energy over the entire fountain height. The code employed is SPARKLE, which uses a spatial discretization of fourth-order accuracy and a third-order Adams-Bashforth time integration scheme. Details of the numerical method used in SPARKLE can be found in \citet{craske2015}.

A Dirichlet boundary condition $w=w_0, b=b_0$ is applied at the source to ensure the uniform source value. We initiate the turbulence by applying an uncorrelated perturbation of 1\% to the velocities in the first cell above the source. The boundary condition at the domain top has a Neumann condition on velocity and buoyancy, and at the domain bottom, the boundary condition is `free slip' with a Neumann condition on buoyancy (except at the source). The four sidewalls of the domain are periodic, raising potential concerns that fluid leaving from one side and re-entering at the opposite can potentially influence the fountain flow --- we mitigate this concern in two ways. First, we ensure that the domain is sufficiently large that the dynamics of the fountain are unaffected by any transport at the domain edges. To do this, we checked a variety of fountain flow statistics for a domain that is about 50\% smaller. We found no substantial differences between the domain in terms of the fountain  (outer boundary) and the exchanges across the internal boundaries (defined later on). These results are presented in appendix \S \ref{App: F1_dom_size}. Second, to avoid any accumulation of buoyancy within the domain affecting our results, a buoyancy `sink' region is set within a thin region (of height $r_0$) at the bottom of the domain and, to avoid affecting the source, only within the region $r \geq 9r_0$ --- within this sink region the fluid buoyancy is gradually adjusted by a `nudging' process towards the environmental value, i.e. the buoyancy is nudged to the zero in this region \citep{Stevens2014}. 

Observations show that the longest timescales in a forced fountain flow are the timescale required to establish a statistically steady state (found here to be approximately $5T_F$), followed by the timescales of the large-scale low-frequency fluctuations near the fountain top. The latter timescale is approximately $2T_F$ \citep{burridge_hunt_2013}. We ran our simulations for a duration of $16T_F$ in order to allow sufficient time for the flow to reach a statistically steady state and then experience a number of the largest scale fluctuations in the region near the top of the fountain, a region which we describe as the `fountain cap'. Instantaneous statistics were obtained at a vertical plane at the centre of the domain with time increments of $0.09T_F$. Azimuthally Reynolds averaged data was computed and stored every $0.14T_F$ once the fountain is in a statistically steady state by partitioning the domain into concentric cylindrical shells and averaging over all cells lying within a given shell \citep{craske2015,Van2016}.

\section{Flow observations and internal structure} \label{Sec:F1_internal_structure}
\subsection{Observations of instantaneous fountain}
\begin{figure}
	\centering{\includegraphics{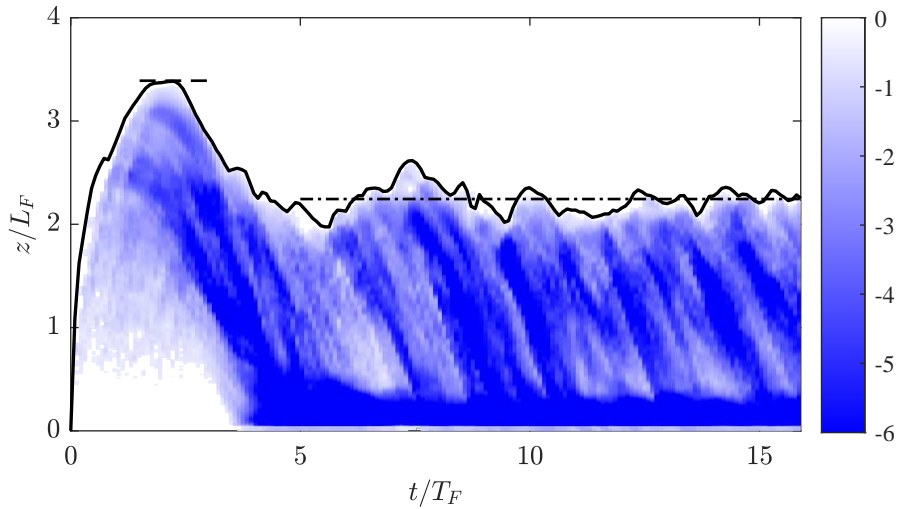}}
	\caption{Time series of normalised local integral buoyancy, $\mathcal{B}(z,t)/\mathcal{B}_0$, within the fountain. The instantaneous fountain height is outlined by the solid line. The horizontal dashed line marks the initial height and the dash-dotted line marks mean steady height. Within the figure, the darker shades of blue represent the greater negative buoyancy.}
	\label{fig:F1_instantaneous_integral_buoyancy} 
\end{figure}
A video showing the fountain from the initial transient state to the statistically steady state can be found within the supplementary material. Figure~\ref{fig:F1_instantaneous_integral_buoyancy} shows a time series of the evolution of the vertical distribution of the horizontally integral buoyancy within the fountain $\mathcal{B}$ (per unit $\pi$), which is defined as
\begin{equation} \label{eq:F1_Bztde}
    \mathcal{B}(z,t) = \frac{1}{\pi}\int_0^{L_x}\int_0^{L_y} b(x,y,z,t) \; \d x \, \d y \approx  \int_{0}^{L_x} b(x,z,t) \left|x-\frac{L_x}{2}\right| \; \d x, 
\end{equation}
for practical reasons (namely, instantaneous data is taken at a central vertical plane), our evaluation comes via the approximation in (\ref{eq:F1_Bztde})).

Fountain flows typically feature a region at the fountain cap which is characterised by the formation and collapse of large-scale structures within the fountain flow as is observed in figure~\ref{fig:F1_instantaneous_integral_buoyancy}. The fountain height is determined by a threshold $ 0.01\,\mathcal{B}(z_b)$, where $\mathcal{B}(z_b)$ is the time-averaged integral buoyancy at a height $z_b/L_F = 1.63$ which we define as the fountain cap base (see \S \ref{sec:F1_internal_structure_of_RAF}). During $t/T_F \lesssim 5.0$ which we regard as the initial transient period, the upflow reaches an initial height at $t/T_F \approx 2.2$, and the downflow is yet to be fully developed until $t/T_F \approx 4.0$. After $t/T_F \approx 5.0$, the fountain can be said to have reached a statistically steady state, and the fountain height fluctuates around a mean steady height. The ratio of initial to steady heights is approximately 3:2, in line with experimental measurements \citep{Turner_1966,burridge_hunt_2012}. Note that the heights referred to here are taken from measurements only in the central vertical plane. Figure~\ref{fig:F1_instantaneous_integral_buoyancy} shows the characteristic (dark blue) inclined stripes that indicate the falling of distinct negatively buoyant fluid structures within the downflow, as previously discussed for fountains in statistically steady state by \citet{burridge_hunt_2013,Mingotti2016}. The thin region, $z/L_F \lesssim 0.05$ in which the local buoyancy is nudged to the environmental value in order to avoid the accumulation of buoyancy in the domain can also be observed.

\subsection{Internal structure of Reynolds averaged fountain} \label{sec:F1_internal_structure_of_RAF}
Once the fountain has reached a statistically steady state, its averaged form is governed by the axisymmetric Reynolds-averaged Navier-Stokes equations in the high-Reynolds number limit, namely
\begin{subequations}
\begin{align} 
\frac{1}{r}\frac{\partial (r\overline{u})}{\partial r}+\frac{\partial \overline{w}}{\partial z}&=0 \, , \label{eq:F1_Reyw}\\
\frac{1}{r}\frac{\partial}{\partial r}(r\overline{u}\,\overline{w}+r\overline{u'w'})+\frac{\partial}{\partial z}({\overline{w}}\,\overline{w}+\overline{w'w'})&=-\frac{\partial \overline{p}}{\partial z} + \overline{b} \, , \label{eq:F1_Reyww}\\
 \frac{1}{r}\frac{\partial}{\partial r}(r\overline{u}\,\overline{b}+r\overline{u'b'})+\frac{\partial}{\partial z}({\overline{w}}\,\overline{b}+\overline{w'b'})&=0, \label{eq:F1_Reywb}
\end{align}
\end{subequations}
where $\overline{\cdot}$ represents the time average (mean): $\overline{b}$ is the mean buoyancy. The mean velocity components ($\overline{u}, \overline{w}$) correspond to the radial direction $r$ and vertical direction $z$ respectively. The mean pressure $\overline{p}$ is the kinematic pressure from which the hydrostatic pressure field resulting from the environmental density has been subtracted.
\begin{figure}
	\centering{\includegraphics{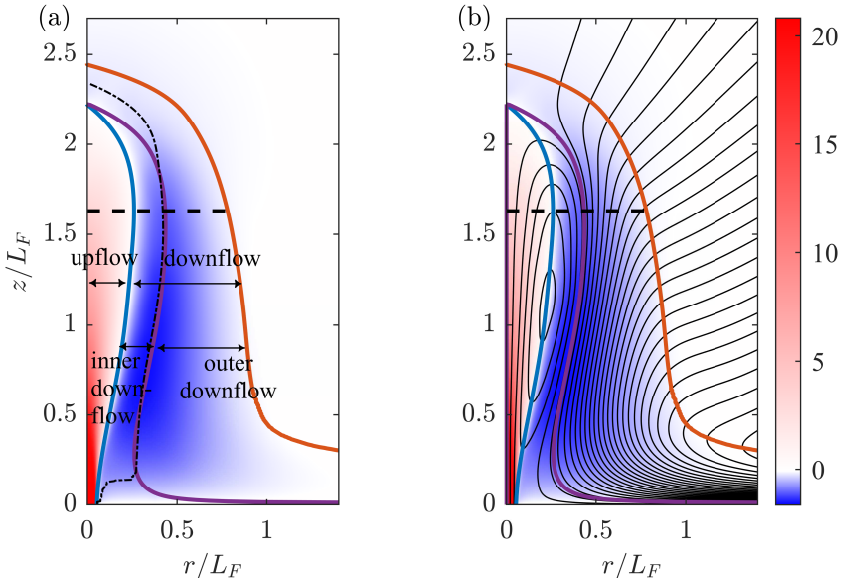}}	
	\caption{The normalised mean vertical velocity $\overline{w}/w_F$, where $w_F = M_0^{1/4}|F_0|^{1/2}$, within the time-averaged fountain marked by colour with regions of: upwards velocities shaded red, downward shaded blue, and zero vertical velocity being white. (a) Highlights the internal structure of the fountain by overlaying the vertical velocities with relevant boundaries: the inner boundary (blue line), outer boundary (red line), and the separatrix (purple line). The thin dash-dotted line marks the loci of points where the maximum downflow vertical velocity occurs. (b) the vertical velocities are overlaid with velocity streamlines. The fountain cap base is marked by the horizontal dashed line.}
	\label{fig:F1_boundaries_and_streamlines}
\end{figure}

Figure~\ref{fig:F1_boundaries_and_streamlines}(a) shows the internal structure of the Reynolds-averaged fountain in its statistically steady state with the shaded colour indicating the mean vertical velocity, $\overline{w}=\overline{w}(r,z)$, at that location. We define the point at which the mean vertical velocity on the centreline first falls to zero to be the `height of the upflow', denoted $z_i$ and we find $z_i/L_F \approx 2.22$. Some works (e.g. BK00) attempted to model fountains only via the application of `plume-like' flow theory, whilst others acknowledged the need to model a `cap' region near the fountain top via distinct considerations \citep[e.g.][and SH14]{McDogall1981}. The fountain cap region can be considered to be the region of flow enclosed by the outer boundary above some height $z_b$, which we term the fountain cap base. The definition used to determine the height of the fountain cap base varies between studies; for example, \citet{McDogall1981} theoretically modelled the cap region as a hemisphere whose radius is half of its base vertical height, SH14 and \citet{Hunt2016} defined the cap base locating at where the local Froude number of upflow is equal to $\sqrt{2}$. In their experimental study of fountains, \citet{Talluru2022} investigated the sensitivity of their findings to the choice cap base definition, dependent on a local Froude number condition (in the upflow); they found that the ratio of the volume flux entrained within the cap, to that in the upflow at the cap base, was insensitive to the location of the cap base when the local Froude number condition was chosen to be in the range 1.2--1.8. The numerical study of \citet{Awin2018} defined the cap base at the location where the radius of upflow is the widest. Specifically for our fountain, the locations of the cap base following each of the three above definitions from the existing literature are all within the region $1.60 \leq z/L_F \leq 1.65$. Therefore, our precise choice of definition is of little significance and, somewhat arbitrarily, we choose to define the fountain cap base as the point of maximum upflow radius, giving, $z_b/L_F = 1.63$. 

 The downflow of the steady fountain and the environment are separated by the outer boundary of radius $r_f=r_f(z)$, this also constitutes the boundary of fountain flow. The outer boundary is defined by the loci of points at which the mean buoyancy $\overline{b} = 0.01 \overline{b}_{cc}$, where $\overline{b}_{cc}=\overline{b}(0, z_b)$ is the centreline mean buoyancy at the fountain cap base level which is about $0.36 b_0$. The sensitivity of our results to the constant within the threshold $\overline{b} = 0.01 \overline{b}_{cc}$, that was used to determine the outer boundary is presented in appendix \S \ref{App: F1_dom_size}; this sensitivity is not deemed significant. The top of the outer boundary (also the final steady height of the entire time-averaged fountain) is located at $z_f/L_F \approx 2.45$, i.e. $z_f/r_0 \approx 2.45Fr_0$, which 
is close to the value $2.46$ reported by \citet{burridge_hunt_2012}, and also agrees well with other experimental literature \citep[see][]{Turner_1966,Mizushina1982,baines1990}. The outer boundary generally increases from the top to bottom as, too, does the area of downflow. We find that the height of the cap region (from cap base to the top) is $z_f-z_b = 0.82L_F$, which comprises about one-third of the total fountain height. We note that the fountain cap region is not our primary focus but we do include some analysis in \S \ref{sec: F1_cap}, and we mark the location of the fountain cap base in all relevant figures for completeness.

\vspace{0.5\baselineskip}
Below the cap, the flow can be divided into two regions separated by an internal boundary, or interface. Herein, we investigate two different definitions of this internal boundary. The first is to decompose the fountain into an upflow region where $\overline{w}$ is positive (shaded red in figure~\ref{fig:F1_boundaries_and_streamlines}) and a downflow region where $\overline{w}$ is negative (shaded blue), that are separated by the inner velocity boundary $r_i$ defined by the loci of points at which $\overline{w} = 0$. The radial position of this inner velocity boundary (hereinafter `inner boundary') $r_i$ increases to a maximum width at $z/L_F \approx 1.63$ (in fact the definition of cap base) and decreases to zero above, at the point at which the mean centreline velocity reduces to zero. Shown in figure~\ref{fig:F1_boundaries_and_streamlines}(b) are the streamlines of the mean velocity, $\{\overline{u},\overline{w}\}$. By construction, the streamlines are horizontal on the inner boundary since the vertical velocity is zero. 

The streamlines in figure~\ref{fig:F1_boundaries_and_streamlines}(b) then provide our second definition of the internal boundary. The streamlines illustrate that the flow being injected from the source is, in the mean, separated from the entrained ambient flow by a streamline that originates from the stagnation point at the top of the inner boundary. This separatrix, which we denote $r_s(z)$, is highlighted by the purple line in figure~\ref{fig:F1_boundaries_and_streamlines}(a) and defines our second internal boundary. The streamlines outside the separatrix show the environmental fluid is entrained into the downflow and eventually leaves the domain from the side. The streamlines cross the outer boundary at some angle to the horizontal indicating that the entrainment velocity has some vertical component. The streamlines inside the separatrix either originate from the source or are closed loops within the upflow, crossing the inner boundary twice --- this highlights that, in the mean, the upflowing core entrains near the source and then detrains fluid into downflow further from the source (above $z/L_F \approx 1.20$). 

By construction, there is no mean exchange across the separatrix, and so the volume flux inside $r_s$ is conserved. Exchange across the separatrix $r_s$ can thus only occur by turbulence. 
By definition, the separatrix intersects the centreline at the height $z=z_i$, i.e. $r_s(z_i) = r_i(z_i) = 0$. The separatrix divides the downflow into an `inner downflow region' (the region between the inner boundary and separatrix), and an `outer downflow region' (the region between the separatrix and outer boundary). The outer downflow region contains all mass (volume) that has been entrained from the environment. These regions and boundaries are all shown in figure~\ref{fig:F1_boundaries_and_streamlines}(a). Except close to the fountain top, the location of the separatrix (purple line) is very similar to that of the vertical velocity maximum (thin black dashed line). This suggests that the fluid beyond the vertical velocity maximum is entrained from the environment.

\begin{figure}
	\centering{\includegraphics{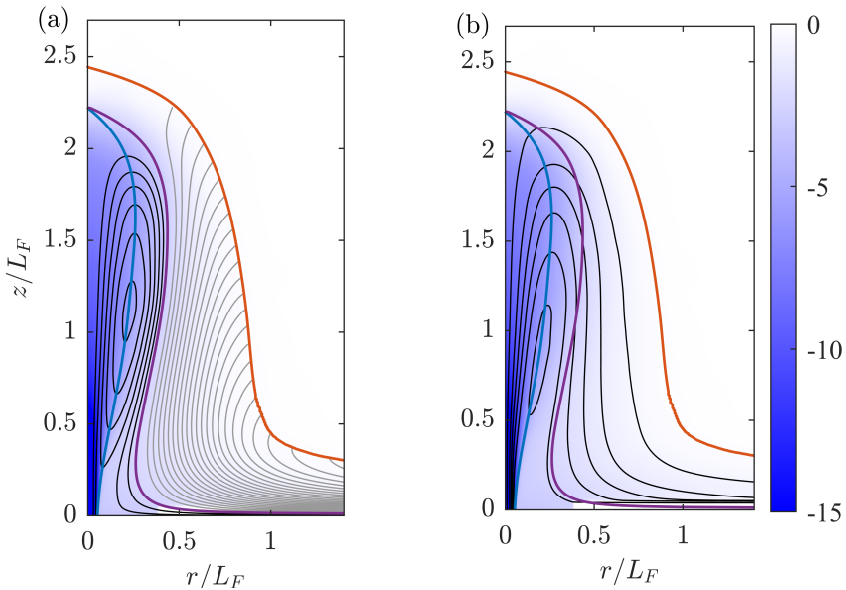}}	
	\caption{The normalised mean buoyancy field $\overline{b}/b_F$, $b_F =M_0^{-5/4}|F_0|^{3/2}$, overlaid with three fountain boundaries presented in figure~\ref{fig:F1_boundaries_and_streamlines} and: (a) field lines of mean buoyancy flux $\{\overline{u}\overline{b},\overline{w}\overline{b}\}$, where the grey lines mark the field lines outside the separatrix; (b) field lines of total buoyancy flux $ \{ \overline{u}\overline{b}+\overline{u'b'}, \overline{w}\overline{b}+\overline{w'b'} \}$.}
	\label{fig:F1_fieldlines}
\end{figure}

\vspace{0.5\baselineskip}
The relevance of the separatrix $r_s$ to fountain flows is highlighted by consideration of the field lines of mean buoyancy flux $\{\overline{u}\overline{b},\overline{w}\overline{b}\}$, and total buoyancy flux $ \{ \overline{u}\overline{b}+\overline{u'b'}, \overline{w}\overline{b}+\overline{w'b'}\}$, plotted in figures~\ref{fig:F1_fieldlines}(a) and \ref{fig:F1_fieldlines}(b), respectively, both with the buoyancy field overlaid. Here we note that, as \eqref{eq:F1_Reywb} demonstrates, only the total buoyancy flux is a conserved quantity. This implies that for the field lines of mean buoyancy, there is a sink-source term and the total buoyancy between the two field lines need not be conserved. As the mean buoyancy flux is the product of the velocity field and the scalar buoyancy field, the field lines of mean buoyancy flux are similar to the streamlines inside the fountain presented in figure~\ref{fig:F1_boundaries_and_streamlines}(b) within the separatrix. As the environment has neutral (zero) buoyancy, the fountain source is the only source of (negative) buoyancy. Therefore it is worth noting from figure~\ref{fig:F1_fieldlines}(a) that: firstly, no field lines exist outside the outer boundary $r_f$; and secondly, the buoyancy injected from the source cannot be transported outside of the separatrix $r_s$ by the mean buoyancy fluxes. However, we can observe there is mean transport of (negative) buoyancy flux within the outer downflow beyond, as indicated by the grey field lines between the purple boundary, $r_s$, and red boundary, $r_f$. Combining observations from figure~\ref{fig:F1_fieldlines}(a) and figure~\ref{fig:F1_fieldlines}(b) highlights that it is only turbulence that transfers the negative buoyancy across $r_s$ into the outer downflow region, i.e. \emph{all} the negative buoyancy within the outer downflow is the result of turbulent exchange across $r_s$. The field lines of total buoyancy flux in figure~\ref{fig:F1_fieldlines}(b) evidence the transfer of buoyancy across the separatrix due to turbulence. Since the total buoyancy flux within the fountain is conserved, the buoyancy flux in the upflow equals the buoyancy flux in the downflow.

Note that the streamlines and field lines indicate that, as expected, the dynamics of the outer downflow below $z/L_F \approx 0.5$ are visibly affected by the (free-slip) bottom boundary, e.g. the flow is forced radially outwards in that region. To ensure the absence of boundary effects in the data presented herein, we therefore only display data from the outer downflow region for $z/L_F > 0.66$ in all further analyses. However, the inner downflow region is not likely to be significantly affected since this fluid is entrained by the upflow. We, therefore, consider our results for the inner downflow region to be valid over the full height.

\section{Integral equations} \label{Sec:F1_equations}
The theoretical work in this section presents exact transport budgets for each flow region. We integrate \eqref{eq:F1_Reyw}-\eqref{eq:F1_Reywb} over the radial coordinate $r$ between two radial locations $r_1(z)$ and $r_2(z)$ and employ the Leibniz integration rule to give conservation equations (per unit $\pi$) for the fluxes of volume, momentum, and buoyancy as
\begin{subequations}
\begin{align}
\label{eq:F1_Qbud}
\frac{\d Q}{\d z} &= q_1-q_2\, ,\\    
\label{eq:F1_Mbud}
\frac{\d (M + M' + P)}{\d z} &= B+ m_1 -m_2\, ,\\
\label{eq:F1_Fbud}
\frac{\d (F+F')}{\d z} &= f_1-f_2\, .
\end{align}
\end{subequations}
The integral quantities, per unit $\pi$, are defined as
\begin{equation} \label{eq:F1_integralquantities}
\begin{aligned}
Q &\equiv 2 \int_{r_1}^{r_2}\overline{w}r\,\d r, & B &\equiv 2\ \int_{r_1}^{r_2}\overline{b}r\,\d r, && \\
M &\equiv 2\ \int_{r_1}^{r_2}\overline{w}^2r\,\d r, & M' &\equiv 2\ \int_{r_1}^{r_2}\overline{w'^2}r\,\d r,
& P &\equiv 2\ \int_{r_1}^{r_2}\overline{p}r\,\d r,\\
F &\equiv 2 \int_{r_1}^{r_2}\overline{w}\overline{b}r\,\d r, & F' & \equiv 2 \int_{r_1}^{r_2}\overline{w' b'}r\,\d r, &&\\
\end{aligned}
\end{equation}
where $B$ is the integral mean buoyancy. The quantities $M$, $M'$ and $P$ correspond to the integral mean momentum flux, turbulent momentum flux and pressure.
Moreover, $F$ and $F'$ denote the integral mean buoyancy flux and turbulent buoyancy flux.
Depending on the region of interest, the inner boundary can take the locations $r_1(z) = \{0, r_i(z), r_s(z) \}$, and the outer boundaries $r_2(z) = \{r_i(z), r_s(z), r_f(z) \}$, with the only condition being that $r_2(z) > r_1(z)$. The perpendicular exchanges of volume, momentum and buoyancy across the boundary $r_j$ (with $j=i,f,\textrm{ or }s$) are defined as 
\begin{subequations}
\label{eq:F1_radialfluxes}
\begin{align}
\label{eq:F1_qj}
    q_j &= 2r\left(\overline{u}-\overline{w}\frac{\d r}{\d z}\right)\Bigg|_{r_j},\\
\label{eq:F1_mj}
    m_j &= 2r \left( \overline{u}\,\overline{w}-{\overline{w}}^2\frac{\d r}{\d z} \right)\Bigg|_{r_j} 
    + 2r\left(
	\overline{u'w'}-\overline{w'^2}\frac{\d r}{\d z}\right)\Bigg|_{r_j} 
	+2r\left( -\overline{p}\frac{\d r}{\d z}\right) \Bigg|_{r_j} \; \textrm{and},\\    
\label{eq:F1_fj}
    f_j &= 2r \left( \overline{u}\,\overline{b}-{\overline{w}\,\overline{b}}\frac{\d r}{\d z} \right)\Bigg|_{r_j} 
    + 2r\left(
	\overline{u'b'}-\overline{w'b'}\frac{\d r}{\d z}\right)\Bigg|_{r_j} \, ,
\end{align}
\end{subequations}
respectively. The terms associated with the radial change, i.e. $ \d r/\d z$, are `Leibniz terms'. As discussed in \S \ref{Sec:F1_introduction}, the perpendicular exchange of volume flux $q_j$ (\ref{eq:F1_qj}) only comprises a mean contribution, and the mean entrainment velocity  $\mybar{V}_g$, see \eqref{eq:F1_V_g}, is clearly recognisable. The term $m_j$, (\ref{eq:F1_mj}), is the perpendicular exchange of momentum flux which is comprised of a mean momentum flux (the first term on the RHS; note that this term is equal to  $2 r \mybar{V}_g \overline{w}$ evaluated at $r_j$) and turbulent momentum flux terms (the second and third terms on the RHS; with the second term equal to $2r \overline{V_g'w'}$ evaluated at $r_j$ and the third term accounting for the effects of pressure which are typically small compared to the second term). The term $f_j$, (\ref{eq:F1_fj}), is referred to as the perpendicular exchange of buoyancy flux which is comprised of one term for mean buoyancy entrainment (the first term on the RHS; equal to  $2 r \mybar{V}_g \, \overline{b}$ evaluated at $r_j$) and one for turbulent buoyancy entrainment (the second term on the RHS; equal to $2r \overline{V_g'b'}$ evaluated at $r_j$).
Whether these exchanges are termed `entrainment' or `detrainment' depends solely on the chosen `region of interest' and the sign of the flux.

In the remainder of this section, the budgets for the upflow, downflow and inner downflow are presented by substituting appropriate boundaries into the equations above. For completeness, we note that since we use a buoyancy that is consistently defined relative to the uniform ambient fluid our buoyancy formulation corresponds to both that termed `BFI' by BK00, which is also that used by SH14.

\subsection{Upflow region}
The upflow budgets can be obtained by integrating from the centreline to the inner boundary, i.e. $r_1(z)= 0$ and $r_2(z) = r_i(z)$. Denoting upflow quantities with subscript `$u$', it immediately follows that (\ref{eq:F1_Qbud}--\ref{eq:F1_Fbud}) for the upflow region become
\begin{subequations}
\label{eq:F1_du}
\begin{align}
\label{eq:F1_dQu}
\frac{\d Q_u}{\d z} &= -q_i\, ,\\    
\label{eq:F1_dMu}
\frac{\d (M_u + M'_u + P_{u})}{\d z} &= B_u  -m_i\, ,\\
\label{eq:F1_dFu}
\frac{\d (F_u+F'_u)}{\d z} &= -f_i\, .
\end{align}
\end{subequations}

By stipulating that $Q_u = r_u^2 w_u$, $M_u = r_u^2 w_u^2$ and $B_u = r_u^2  b_u$, the characteristic (top-hat) upflow vertical velocity $w_u$, buoyancy $b_u$, and radius $r_u$ are defined
\begin{equation} \label{eq.F1_charu}
 w_{u} \equiv \frac{M_u}{Q_u}, \quad b_{u} \equiv \frac{B_u M_u}{Q_u^2}, \; \textrm{and} \quad r_{u} \equiv \frac{Q_u}{M_u^{1/2}}.
\end{equation}
These natural characteristic scales will be used to investigate the entrainment coefficients in section \ref{sec:F1_coefficients}.

\subsection{Downflow region} \label{sec:F1_df_eq}

The downflow budget (i.e. including both the inner and outer downflow), can be obtained by integrating from the inner boundary to the fountain edge, i.e. $r_1(z)= r_i(z)$ and $r_2(z) = r_f(z)$. As the flow direction is opposite to the upflow, we define a downward coordinate $z_d$. A transformation $z\rightarrow -z$ implies that $w \rightarrow -w$, $b \rightarrow -b$ in order to achieve invariance of the equations (\ref{eq:F1_Reyw}--\ref{eq:F1_Reywb}). From the definitions, \eqref{eq:F1_integralquantities}, it now follows that the integral downflow quantities are given by $Q_d = -Q$,  $M_d = M$,   $F_d = F$,  $B_d = -B$ and $ {\d}/{\d z_d} = -{\d}/{\d z}$, with the transformation of the turbulent quantities following an equivalent form.
Substitution into (\ref{eq:F1_Qbud})--(\ref{eq:F1_Fbud}) results in
\begin{subequations}
\label{eq:F1_dd}
\begin{align}
\label{eq:F1_dQd}
\frac{\d Q_d}{\d z_d} &= q_i-q_f,\\   
\label{eq:F1_dMd}
\frac{\d (M_d + M'_d + P_{d})}{\d z_d} &= B_d + m_f -m_i,\\
\label{eq:F1_dFd}
\frac{\d (F_d+F'_d)}{\d z_d} &= f_f-f_i.
\end{align}
\end{subequations}

The radial coordinate remains unchanged in this transformation. Therefore, a positive volume flux $q_f$ contributes negatively to the downflow budget. Note that at the outer boundary, the term `$m_f$' can be interpreted as $-(-m_f)$, where $-m_f$ is an outwards flux of downward momentum (consider (\ref{eq:F1_mj}) at $r_j = r_f$ under coordinate $z_d$). 
Similar understanding holds true for the buoyancy flux term $f_f$ in (\ref{eq:F1_dFd}). At the inner boundary, the exchange of volume flux, $q_i$, contributes oppositely to the upflow and downflow (\ref{eq:F1_dQu},\ref{eq:F1_dQd}) as the volume flux exchanged out of the upflow will increase the volume flux in the downflow and \textit{vice versa}. However, in terms of the upflow, $m_i$ represents an outward flux of upward momentum which reduces the upflow momentum flux, so that a negative sign is correctly associated with $m_i$ in (\ref{eq:F1_dMu}); with respect to the downflow and the coordinate $z_d$, $-m_i$ denotes a flux of downward momentum into the downflow (radially outward) at the inner boundary, this contributes positively to the downflow momentum. Hence the sign convention in (\ref{eq:F1_dMu}) and (\ref{eq:F1_dMd}), with equivalence, for the term $f_i$ in (\ref{eq:F1_dFu}) and (\ref{eq:F1_dFd}).

The downflow region has a characteristic annular cross-section area, which can be denoted (per unit $\pi$) as $r_d^2 - r_u^2$ (BK00); thus, the fluxes can be written as
$Q_d = (r_d^2-r_u^2) w_d$, $M_d = (r_d^2-r_u^2) w_d^2$, $B_d = (r_d^2-r_u^2) b_d$ and it follows that the natural characteristic scales are
\begin{equation} \label{eq.F1_chard}
 w_{d} \equiv \frac{M_d}{Q_d}, \quad b_{d} \equiv \frac{B_d M_d}{Q_d^2},\;\textrm{and} \quad r_{d} \equiv \sqrt{\frac{Q_u^2}{M_u}+ \frac{Q_d^2}{M_d}}.
\end{equation}

\subsection{Inner downflow region}
The transport budgets for the inner downflow region can be obtained by integrating from the inner boundary to the separatrix, i.e. $r_1(z)= r_i(z)$ and $r_2(z) = r_s(z)$. As the flow in this region is by definition downward, the transformation applied to \S \ref{sec:F1_df_eq} is required. Denoting inner downflow quantities with subscript `$id$', we obtain
\begin{subequations}
\label{eq:F1_idd}
\begin{align}
\label{eq:F1_dQid}
\frac{\d Q_{id}}{\d z_d} &= q_i-q_s,\\    
\label{eq:F1_dMid}
\frac{\d (M_{id}+ M'_{id} + P_{id})}{\d z_d} &= B_{id} + m_s -m_i,\\
\label{eq:F1_dFid}
\frac{\d (F_{id}+F'_{id})}{\d z_d} &= f_s-f_i.
\end{align}
\end{subequations}

\section{Integral quantities} \label{Sec:F1_Integral_quantities}
\begin{figure}
	\centering{\includegraphics{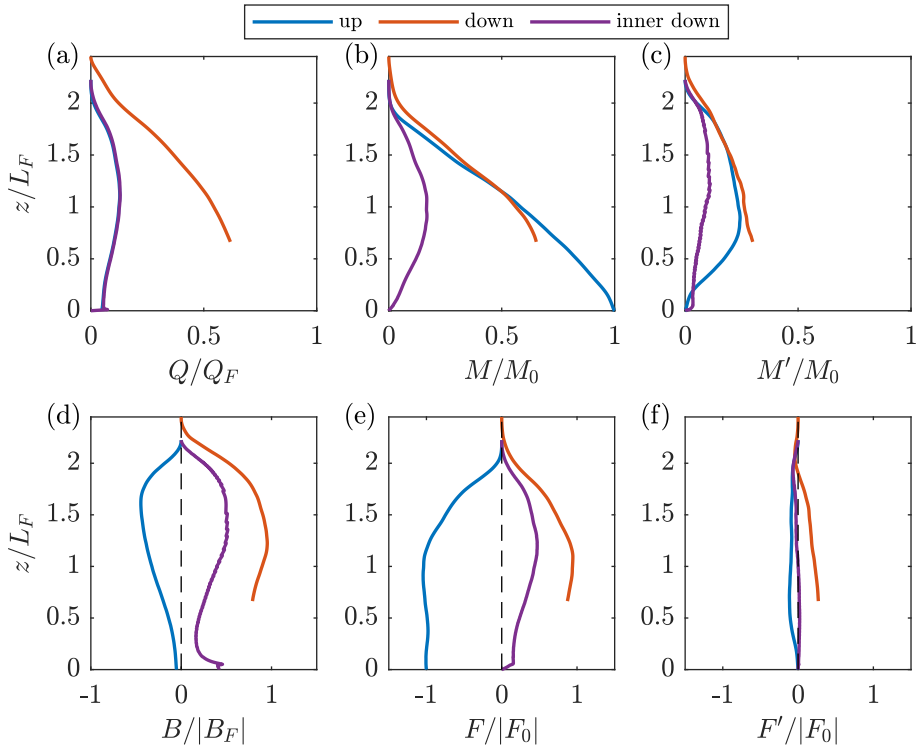}}	
	\caption{Integral quantities for the upflow, downflow and inner downflow region, each normalised by the relevant source scales. The vertical dashed marks fluxes of zero where appropriate.}
	\label{fig:F1_fluxes}
\end{figure}

The vertical variation in the normalised integral quantities in the upflow, downflow, and inner downflow (the portion of the downflow that falls within the separatrix) are shown in figure~\ref{fig:F1_fluxes}. Buoyancy fluxes and integral buoyancy are normalised by absolute values of $B_F$ and $F_0$, respectively, in order to accurately reflect the opposing or assisting action of the buoyancy force on each of the flows. We choose to discuss any flow regions in the direction of their flow, i.e. we discuss any downward flowing regions from top to bottom (that is in the direction of the downwards coordinate $z_d$). 

Figure~\ref{fig:F1_fluxes}(a) shows that below $z/L_F \approx 1.11$, the upflow volume flux increases, suggesting a net volume entrainment into the upflow in this region.
Above $z/L_F \approx 1.11$ the volume flux in the upflow decreases due to the volume detrainment into the downflow. The figure also shows that the volume flux in the upflow is identical to that within the inner downflow; this confirms that, by construction, the volume flux is conserved within the separatrix.
The volume flux in the downflow increases from the top of the fountain showing net volume entrainment by the downflow. Noting that the downflow consists of both inner- and outer-downflow regions, the difference between the downflow and the inner downflow, reassuringly, confirms that fluid from the environment is entrained into the fountain outer downflow over the full height of the outer boundary. Careful observation of the streamlines in figure~\ref{fig:F1_boundaries_and_streamlines}(a) confirms these above observations.

Figure~\ref{fig:F1_fluxes}(b) plots the normalised vertical evolution of mean vertical momentum fluxes. The mean momentum flux in the upflow, $M_u$, decreases, to zero at the top of the upflow $z_i$, as the buoyancy acts to decelerate the flow; whilst this buoyancy then acts to enhance the mean momentum flux in the downflow $M_d$. The momentum flux within the inner downflow $M_{id}$ is also shown to increase from the top of the fountain until the height $z/L_F \approx 0.88$, below which $M_{id}$ then decreases back towards zero. The magnitudes of the mean momentum flux $M_u$ and $M_d$ are, perhaps surprisingly, approximately equal (despite the significant differences between the characteristic velocity and radial scales within the two flows); this suggests that, broadly speaking, fountain fluid roughly attains the same momentum flux on the way back down that it originally had on the way up.

Figure~\ref{fig:F1_fluxes}(c) plots the normalised turbulent vertical momentum flux. The maximum turbulent momentum flux in upflow $M'_u$ occurs relatively low down within the fountain (about $z/L_F \approx 0.88$, i.e. about a third of the height of the fountain), the reason for this is unclear. The turbulent momentum flux in the downflow $M'_d$ continuously increases from the fountain top, in line with the suggestion that, since the buoyancy increases the mean momentum flux, the turbulent momentum flux broadly scales with the mean momentum flux. Again, for heights $z/L_F \lesssim 1.92$, there is a surprisingly close agreement between the magnitude of the turbulent momentum fluxes in the upflow and downflow, perhaps suggesting that both these turbulent fluxes might be predominantly driven by the mean momentum fluxes over much of the fountain height. In-depth analysis of the data in figures~\ref{fig:F1_fluxes}(b) and ~\ref{fig:F1_fluxes}(c) show that the ratio of turbulent and mean momentum flux $M'/M$ is greatest near the fountain top, suggesting analysis of the dynamics via mean flow statistics might be insufficient in this region.

Figure~\ref{fig:F1_fluxes}(d) shows that the integral buoyancy in the upflow $B_u$ is negative, while in the downflow $B_d$ is positive. Combining (\ref{eq:F1_dMu}) and (\ref{eq:F1_dMd}), confirms that the integral buoyancy is, in fact, a sink for $M_u$ but a source for $M_d$. The magnitudes of all the three flow regions peak near $z/L_F \approx 1.50$. The inner downflow region contains nearly half of the integral buoyancy in the downflow, evidence that the buoyancy within the inner downflow is more concentrated than the outer downflow (i.e., the inner downflow is denser than the outer downflow).

Figure~\ref{fig:F1_fluxes}(e) and (f) show the mean and turbulent vertical buoyancy fluxes respectively. Unlike the momentum fluxes, the turbulent buoyancy fluxes are much smaller than the mean in all flow regions, suggesting that the turbulence affects the buoyancy fluxes less than it does the momentum fluxes. The upflow mean buoyancy flux $F_u$ is negative due to its opposing buoyancy; it is noteworthy that for $z/L_F \lesssim 1.20$, the mean buoyancy flux in the upflow is approximately constant suggesting that in this region, whilst the integral buoyancy within the upflow is increasing (figure~\ref{fig:F1_fluxes}(d)), a particular interplay between radial growth, upflow velocity, and entrainment from the downflow, must exist to maintain this approximately constant flux which then breaks down at greater heights. When looking at the mean quantities, the inner downflow buoyancy flux is approximately half that of the total downflow. However, the turbulent buoyancy flux in the inner downflow region is a much smaller fraction of the total, implying that most of the downflowing turbulent buoyancy flux is transported in the outer downflow region. Finally, figure~\ref{fig:F1_fluxes}(e) and (f) show that, although relatively weak, the turbulent buoyancy fluxes act in the same direction as the mean buoyancy fluxes within both the upflow and downflow.

\section{Entrainment} \label{Sec:F1_transfer}
The rich variety of exchanges, herein `entrainment', between the fountain upflow, the returning downflow, and the fountain cap is crucial in determining the bulk dynamics of fountain flows, including the interactions with the environment. We now examine these in detail.
\begin{figure}
	\centering{\includegraphics{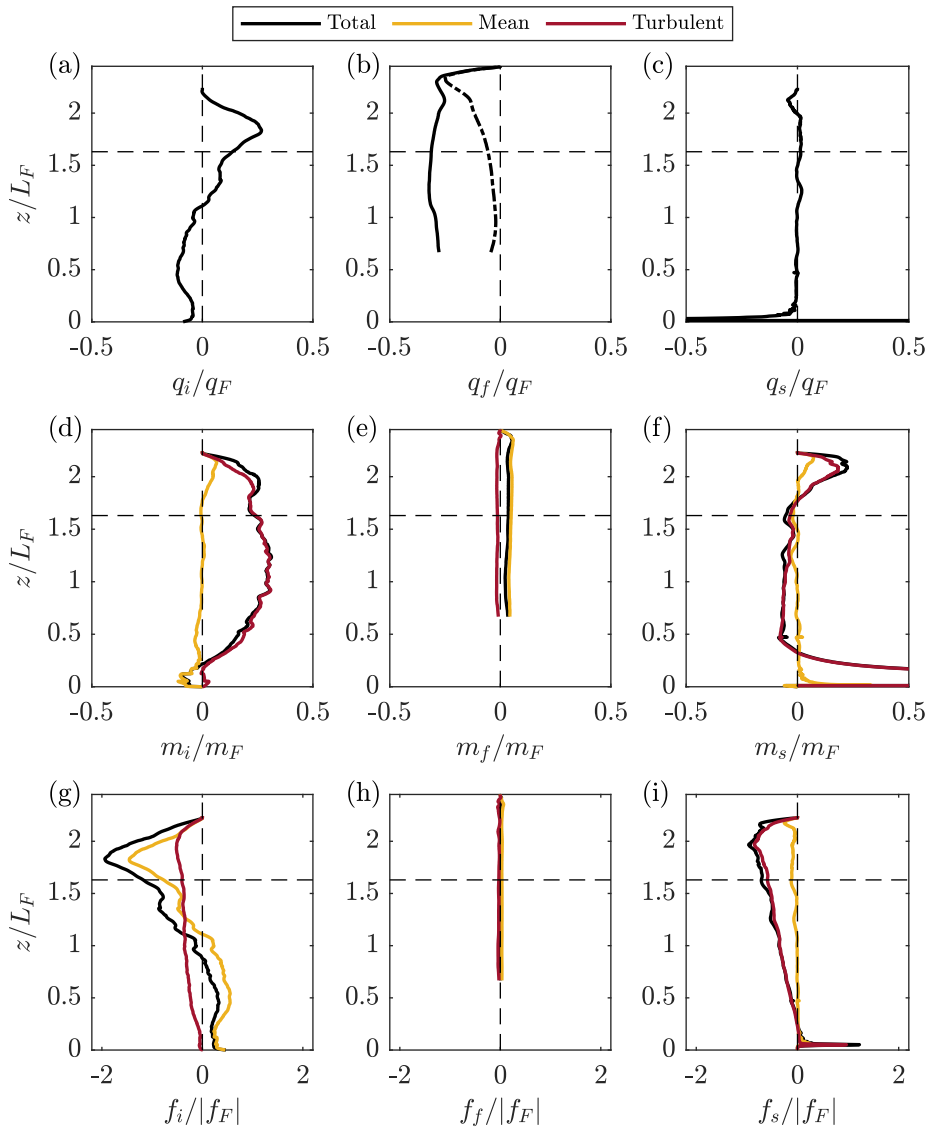}}
\caption{The vertical variation of the normalised (a--c) volume entrainment, (d--f) total momentum entrainment and (g--i) total buoyancy entrainment at the (left column) inner boundary, (middle column)  outer boundary and (right column) the separatrix respectively. The mean and turbulent components of momentum and buoyancy entrainment are included. The total volume entrainment is also the mean. The dash-dotted line in (b) shows the vertical component of mean volume entrainment. The horizontal dashed line marks the fountain cap base. The vertical dashed line marks the line of zero exchange.}
	\label{fig:F1_entrainment}
\end{figure}
\subsection{Vertical evolution of entrainment} \label{sec:F1_ent}

Figure~\ref{fig:F1_entrainment} shows the entrainment terms, see (\ref{eq:F1_radialfluxes}), at each of the defined internal boundaries $r_i$, $r_s$ and the outer boundary $r_f$. As expected all entrainment terms approach zero at the top of their respective boundary (i.e. where these boundaries also approach $r=0$).

Results for the volume exchange at the inner boundary (the term $q_i$, figure~\ref{fig:F1_entrainment}(a)) are reassuringly supportive of the observation of the fluxes in figure~\ref{fig:F1_fluxes}(a). We note that for the upflow region, there is a change from net entrainment to net detrainment at $z/L_F \approx 1.11$, where the local Froude number of upflow is $2.87$, about twice of the value at the fountain cap base. By construction, the mean momentum entrainment is close to zero at the inner boundary (figure~\ref{fig:F1_entrainment}(d)), but the momentum exchange associated with turbulence by the upflow is significant over much of the fountain height --- this acts to reduce both the momentum fluxes of the upflow (\ref{eq:F1_dMu}) and downflow (\ref{eq:F1_dMd}), either by entrainment of downward momentum fluctuations into, or detrainment of upward momentum fluctuations from the upflow; which of these is dominant could be rigorously determined by investigations of the instantaneous transports, for example, via a quadrant plot but this is beyond the scope of the current study.
Alternatively, in terms of the turbulent stress $-\overline{u'w'}$ on the boundary, the interpretation of the reduction of upflow momentum flux is that it experiences substantial shear by the downflow, thereby decelerating the upflow. As the upflow and downflow are opposite to one another, the shear will also decelerate the downflow.

The buoyancy exchange at the inner boundary (figure~\ref{fig:F1_entrainment}(g)) exhibits significant contributions from both the mean flow and turbulence. The mean component of $f_i$ exhibits a similar variation to that of $q_i$ (figure~\ref{fig:F1_entrainment}(a)), consistent with the view that buoyancy can be both entrained and detrained by the mean flow at different heights according to $\mybar{V}_g \overline{b}$. By contrast, the turbulent component of $f_i$ is negative at all heights; suggesting that with respect to the upflow, there is either detrainment of negative buoyancy fluctuations, or entrainment of positive buoyancy fluctuations. 

\begin{figure}
    \centering
    \includegraphics[width=13cm]{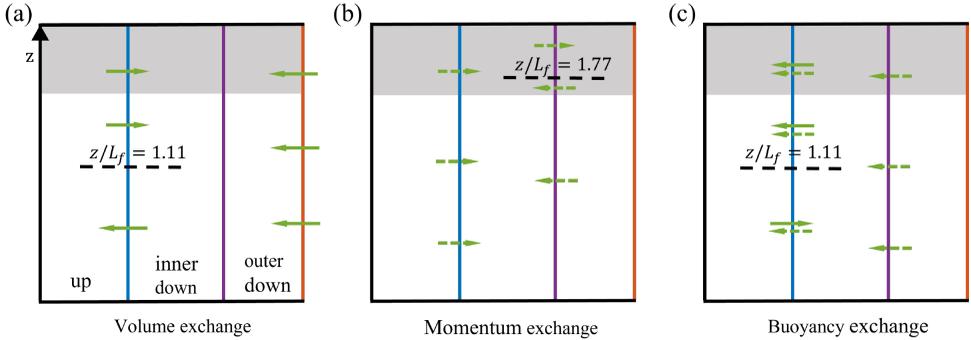}
    \caption{Schematic illustration depicting the sign of the exchange fluxes across the identified boundaries: right being positive and left being negative of the exchanges across the relevant boundaries within the fountain: (a) volume exchange, (b) momentum exchange and (c) buoyancy exchange. From left to right, each pane contains vertical lines: black marks the fountain's centre line, blue illustrates the inner boundary, purple the separatrix, and orange the outer boundary. Green horizontal solid arrows represent the fluxes associated with the mean flow, while green dashed arrows represent those associated with turbulence. The dashed black lines mark any height at which the exchange changes sign. The grey areas represent the fountain cap region. The vertical coordinate is drawn approximately to scale, but the scale of the arrows itself does not represent the magnitude of the entrainment exchanges.}
    \label{fig:F1_ent_sketch}
\end{figure}

A schematic illustration of the net flux $q_j$, $m_j$ and $f_j$ (with $j=i,f,\textrm{ or }s$) shown in figure \ref{fig:F1_ent_sketch} is intended to clarify the sign of the exchanges. During the following discussion, this sketch is intended to act as a useful visual reference of the exchanges. The arrows indicate the sign of the exchanges of volume, momentum and buoyancy between the vertical fluxes of relevant flow regions, at various heights within the fountain. Taking the above inner boundary as an example, in figure~\ref{fig:F1_ent_sketch}(a), the volume exchange across the inner boundary (blue line) $q_i$ below $z/L_f = 1.11$ has a leftward arrow which denotes a negative flux (in this case radially inward). Combined with \eqref{eq:F1_dQu}, this illustrates that $q_i$ increases the volume flux in the upflow in that region. 

At the outer boundary (figure~\ref{fig:F1_entrainment}(b)), there is mean volume entrainment from the environment into the downflow. Inspired by the streamline patterns shown in figure~\ref{fig:F1_boundaries_and_streamlines}(b), we include results for the vertical component of volume entrainment at the outer boundary within figure~\ref{fig:F1_entrainment}(b); for which the contribution is significant within the region $z/L_F \gtrsim 1.63$, i.e. within the fountain cap region. Our results highlight that the volume entrainment into the fountain from the environment is generally greater than the volume exchange within the fountain (see figure~\ref{fig:F1_entrainment}(a)). Figure~\ref{fig:F1_entrainment}(e) shows there is very low momentum entrainment into the downflow at the outer boundary due to low velocities and levels of turbulence beyond; reassuringly, the buoyancy entrainment at the outer boundary, $f_f$, is shown to be negligible (figure~\ref{fig:F1_entrainment}(h)).

Along the separatrix there is, by construction, no volume exchange associated with the mean flow (see figures~\ref{fig:F1_entrainment}(c), (f), and (i)). For $z/L_F \gtrsim 1.77$, $m_s$ is positive, becoming negative at lower heights and lessening the inner downflow momentum flux (\ref{eq:F1_dMid}). Despite the separatrix being at larger radial locations, the magnitude of turbulent momentum exchange there (figure~\ref{fig:F1_entrainment}(f)), is still markedly smaller than that at the inner boundary (figure~\ref{fig:F1_entrainment}(d)), suggesting weaker turbulence at the separatrix than at the inner boundary. Figure~\ref{fig:F1_entrainment}(i) shows that $f_s$ acts to reduce the buoyancy flux in the inner downflow. Note that the figure shows $f_s$ exhibits a linear decrease with the height up to $z/L_F \approx 1.95$, which by adding the integral buoyancy conservation equation \eqref{eq:F1_dFu} and \eqref{eq:F1_dFid}, implying that the total vertical buoyancy flux $F+F'$ of the fountain inside the separatrix grows quadratically with the height (for which, in figure \ref{fig:F1_fluxes}(e) one would need to combine the mean buoyancy flux in the upflow, blue line, with that of the inner downflow, purple line).
   
\subsection{Integral entrainment in the fountain cap region} \label{sec: F1_cap}

As one might expect, the vertical evolution of entrainment of the various quantities (figure~\ref{fig:F1_entrainment}) shows significant changes within the fountain cap region, e.g. sharp peaks in volume and buoyancy exchanges, and the momentum exchange $m_s$ changes the sign around the cap base (figure \ref{fig:F1_entrainment}(f)). Note that in this region, \citet{williamson_armfield_lin_2011} suggested that the momentum transport is significantly affected by the pressure gradient, and is also the region dominated by low-frequency fluctuations that complicate the flow dynamics \citep[e.g.][]{williamson_armfield_lin_2011,burridge_hunt_2013}. In this subsection, we focus on fluxes apparent upon integrating the entrainment over the height of this fountain cap region.

The integral entrainment across the boundaries over the fountain cap is
\begin{equation}
        Q_{cap,j} = \int_{z_b}^{z_t} q_j \d z, \quad M_{cap,j} = \int_{z_b}^{z_t} m_j \d z, \quad F_{cap,j} = \int_{z_b}^{z_t} f_j \d z,
\label{eq:F1_capintegral}
\end{equation}
where $z_t=z_i$ for inner boundary and separatrix, while $z_t=z_f$ for outer boundary. These quantities show the volume, momentum and buoyancy exchanges, respectively, across the corresponding boundary integrated over the cap region. The values are presented in table~\ref{table:t1} with two normalisations: firstly, by the relevant forced fountain scales, e.g., $Q_F$, $M_0$, and $|F_0|$, and secondly, by the upflowing fluxes at the cap base, namely,
\begin{align}
     Q_{ub} = Q_u(z_{b}) & \approx 0.09Q_F, \\
     M_{ub} = M_u(z_{b}) + M'_u(z_{b}) & \approx 0.36M_0\; \textrm{and}, \\
     F_{ub} = F_u(z_{b}) + F'_u(z_{b}) & \approx -0.71|F_0|.
\end{align}
These upflowing fluxes at the cap base highlight that around one-third of the source momentum flux reaches the cap region; and nearly three-quarters of the fountain's total buoyancy flux passes through the cap region. These facts, combined with the results in table~\ref{table:t1}, highlight the importance of the cap region for the dynamics of fountains. Figure \ref{fig:F1_ent_sketch} is intended to aid interpretation of table~\ref{table:t1} by indicating the directions of $Q_{cap,j}$, $M_{cap,j}$ and $F_{cap,j}$. 

\begin{table} 
  \begin{center}
\def~{\hphantom{0}}
  \begin{tabular}{lcccccc}    
                              & $\displaystyle \frac{Q_{cap,j}}{Q_{F}}$ & $\displaystyle\frac{Q_{cap,j}}{Q_{ub}}$ & $\displaystyle\frac{M_{cap,j}}{M_{0}}$ & $\displaystyle\frac{M_{cap,j}}{M_{ub}}$ & $\displaystyle\frac{F_{cap,j}}{|F_{0}|}$  & $\displaystyle\frac{F_{cap,j}}{|F_{ub}|}$
                              \vspace{0.15cm}
                              \\[6pt]
     Fluxes across inner boundary, $r_i$     &   $0.09$ &   $1.00$       &  $0.12$  &  $0.34$   & $-0.75$   & $-1.00$ \\
     Fluxes across separatrix, $r_s$     &   $0.00$ &   $0.00$            &  $0.05$ &  $0.14$   & $-0.45$   & $-0.60$  \\
     Fluxes across outer boundary, $r_f$     & $-0.22$   & $-2.40$    &  $0.02$  &  $0.05$   &  $0.00$  &  $0.00$      \\
  \end{tabular}
    \caption{Integral volume, momentum and buoyancy entrainment across the boundaries within the fountain cap region, normalisation both by the relevant forced fountain scales and by the upflowing fluxes across at the cap base are presented for convenience.}  
    \label{table:t1}
  \end{center}
\end{table}
Table~\ref{table:t1} shows that the integral volume entrainment by the cap is $Q_{cap,f} = -0.22 Q_F$, combining this with the results for the bulk fountain entrainment \citep[i.e. that the total entrainment flux for these highly forced fountains is approximately $0.77Q_F$][]{burridge_hunt_2016}, would suggest that more than a quarter of all fountain's volume entrainment occurs within the cap (we note that our bottom boundary condition prohibits us making our own estimates of the bulk volume entrainment by the fountain). Meanwhile, $Q_{cap,f} = -2.40 Q_{ub}$ shows that the integral volume entrainment from the environment is about two and a half times larger than that injected at the cap base, agreeing with the data taken from \citet{Awin2018} and from \citet{ Talluru2022} for a turbulent forced fountain.

Considering the exchanges across the inner boundary within the cap region, we note that $Q_{cap,i}/Q_{ub} = 1.00$ confirms that, as expected, all upcoming volume flux is detrained into the downflow region. The momentum flux detrained out of upflow, $M_{cap,i}$, is approximately 35\% of the upflowing momentum flux, suggesting around two-thirds of vertical momentum flux that enters the cap is acted on by the pressure gradient and opposing buoyancy. The integral buoyancy detrainment $F_{cap,i}/|F_{ub}| = -1.00$ shows the buoyancy flux injected into the cap is all detrained into the downflow; furthermore, more than half of this is then detrained from the inner downflow to the outer downflow ($F_{cap,s}/|F_{ub}| = -0.60$).

\subsection{Entrainment coefficients} \label{sec:F1_coefficients}

In this section, we calculate the entrainment coefficients from our DNS data, and compare them to those employed in other studies, in particular the fountain model BK00. Classically, entrainment coefficients have been used to represent the rate of dilution of a flow; this approach has proved valuable to the modelling of a broad set of classes of turbulent free-shear flows. This concept was popularised by \citet{Morton1956} (referred to as the MTT entrainment model), who successfully characterised the bulk entrainment of mass by a plume by relating the fluid entrainment velocity across the plume edge to local characteristic scales via a constant entrainment coefficient. The application of this concept to bulk-averaged models of fountains came somewhat later \citep[][]{McDogall1981} and continues to be developed, e.g. BK00 and SH14. In the most general case presented to date, these apply the entrainment coefficients to estimate the mass, momentum and buoyancy exchanges of the upflow and the downflow. Typically, the entrainment coefficients in models were either assumed constant with values determined by analogy to a more simple canonical flow \citep[e.g. a jet and or a line plume, as in the case of][BK00]{McDogall1981}, or approximated empirically based on either experimental or numerical studies of fountains (e.g. SH14). 

Notably, BK00 represented the fountain upflow and downflow as two separate flows: an upward negatively buoyant jet, and a downward (annular) line plume, respectively; which both entrain via their boundaries. As a result, unlike our Reynolds-averaged statistics, they parameterised two-way entrainment at the inner boundary: the upflow entrains from the downflow with a velocity $\omega_{i}$ while the downflow entrains from the upflow with a velocity $\omega_{d}$. Meanwhile, the downflow also entrains from the environment via the outer boundary with a velocity $\omega_{f}$, see figure \ref{fig:BKmodel}. DNS Simulations, or high fidelity experiments, could enable conditionally sampled data to directly inform such a parameterisation (e.g. to parameterise simultaneous entrainment and detrainment of the upflow) should use of such a parameterisation be deemed necessary. In order that readers can check consistency with relative ease, the BK00 model and governing equations are reproduced in appendix \S \ref{App: F1_BKeq} using the notation and coordinate system presented herein.

BK00 considered two formulations of body force acting on the upflow, and two formulations of velocity scale characteristic of entrainment; in combination, this leads to four different cases to consider. In terms of the body force, the first formulation (BF\RNC 1) assumed the buoyancy force acting on the upflow depends directly on the density of the environment, that is in our notation $B_u$ as evident in (\ref{eq:F1_dMu}) and (\ref{eq:F1_BKMu}). The second, BF\RNC 2, assumed the buoyancy force is relative to the local density difference between the upflow and downflow, that is in our notation $B_u+B_d$. In agreement with SH14, we assert that BF\RNC 1 is a uniquely appropriate formulation for the body force; we, therefore, consider only BF\RNC 1. In terms of the characteristic entrainment velocity, the first formulation considered by BK00, which we denote `EI', assumed the mean entrainment velocity should be proportional to the difference in velocity between the upflow and downflow, so that the three entrainment velocities in BK00 can be written as
\begin{equation} \label{eq: F1_omega_I}
-\omega_{i}^I = \alpha_{i_{00}} (w_u+w_d)\, , \quad \omega_{d}^I = \alpha_{f_{00}} w_d \, , \quad -\omega_{f}^I = \alpha_{f_{00}} w_d \, ,
\end{equation}
where, $\alpha_i$ and $\alpha_f$ are the mean volume entrainment coefficient of the upflow and downflow, respectively, the subscript `$00$' denotes the variables as used in BK00, the superscript `I' denotes the entrainment formulation EI. The negative sign in $\omega_{i}$ and $\omega_{f}$ indicates entrainment velocity into upflow and downflow (see appendix \S \ref{App: F1_BKeq} for details). Note that BK00 parameterised the downflow as having the same entrainment coefficient $\alpha_{f_{00}}$ at both boundaries --- the physical reasoning for this remains unclear. 

Neglecting all the turbulent and Leibniz terms, and taking entrainment parameterisation EI, (\ref{eq:F1_dQu}) becomes
\begin{equation} \label{eq:F1_alpha_iI}
q_i = 2\alpha_i^I r_u (w_u+w_d) \, , \quad q_f =  2\alpha_f^I r_d w_d\, ,
\end{equation}
where $\alpha_i^I$ and $\alpha_f^I$ are the entrainment coefficients for the upflow and downflow, respectively, that are evident from our DNS data under entrainment formulation EI. When applying entrainment formulations presented here, it should be clarified that BK00 prescribed the same constant entrainment coefficients, $\alpha_{i_{00}}$ and $\alpha_{f_{00}}$, and then applied two formulations to calculate entertainment velocities apparent within their fountain model; for our DNS data we measured the actual fluxes entrained and then directly determined the relevant entrainment coefficients. Substituting (\ref{eq: F1_omega_I}) into (\ref{eq:F1_BKQu}) and combining with (\ref{eq:F1_dQu}), provides the relationships between our entrainment coefficients and those of BK00, under formation EI, as
\begin{equation}
\label{eq:F1_EI}
    \alpha_i^I = \alpha_{i_{00}} - \alpha_{f_{00}} \frac{w_d}{w_u+w_d} \, , \textrm{ and } \; \; \alpha_f^I = \alpha_f = \alpha_{f_{00}} \;.
\end{equation}

In their second entrainment formulation `EII', BK00 assumed the mean entrainment velocity is related to a characteristic velocity depending only on the relevant fraction of the total shear, i.e. the sum of the magnitude of velocity on either side of the interface. Thus the entrainment equations of BK00 under EII are
\begin{equation} \label{eq: F1_omega_II}
-\omega_{i}^{\RNC 2} = \alpha_{i_{00}} w_u\, , \quad \omega_{d}^{II} = \alpha_{f_{00}} w_d \, , \quad -\omega_{f}^{II} = \alpha_{f_{00}} w_d \; ,
\end{equation}
with the relevant equations for our data being 
\begin{equation} \label{eq:F1_alpha_iII}
q_i = 2\alpha_i^{II} r_u w_u\, , \quad q_f =  2\alpha_f^{II} r_d w_d\, .
\end{equation}
Therefore, the relationships between our entrainment coefficients and those of BK00, under formation EII, become
\begin{equation}
\label{eq:F1_EII}
    \alpha_i^{II} = \alpha_{i_{00}} - \alpha_{f_{00}} \frac{w_d}{w_u} \, , \textrm{ and } \; \; \alpha_f^{II} = \alpha_f = \alpha_{f_{00}} \;.
\end{equation}
We note that, BK00 assumed constant entrainment coefficients throughout the height of the fountain and calibrated their model (against a single statistic of fountain behaviour; namely, the fountain rise height), for the four combinations of body forces and entrainment formulations, always taking the same values for each of their two entrainment coefficients, specifically $\alpha_{i_{00}} = 0.085$ and $\alpha_{f_{00}} = 0.147$ --- for this reason our notation does not distinguish between entrainment formulations when denoting the coefficients of the BK00 model.

The other fountain model worthy of consideration is that of SH14; they assumed that only entrainment by the upflow occurred at the inner boundary (i.e. only inward exchanges over the full height of their upflow $z/L_F \leq 1.62$). With this assumption, (\ref{eq:F1_EI}) and (\ref{eq:F1_EII}) reduce to $\alpha_i^{I} = \alpha_i^{II} = \alpha_{i_{00}}$. SH14 took the inner boundary entrainment coefficient to be $\alpha_i =0.06$, and $\alpha_f = 0.15$ for the entrainment coefficient at downflow at outer boundary \citep[these values being loosely based on those reported in the studies of][]{williamson_armfield_lin_2011, burridge_hunt_2013} --- the precedents set by SH14's model were followed by \citet{debugne_hunt_2016}. 

 \begin{figure}
     \centering
     \includegraphics{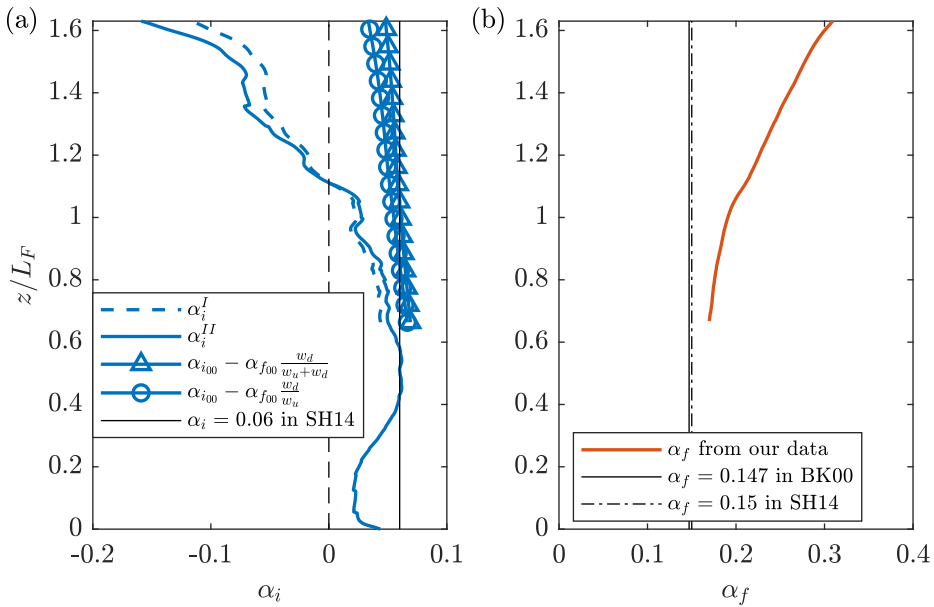}
     \caption{The vertical evolution of entrainment coefficients both calculated based on Reynolds averaging our DNS data ($\alpha_{i}^{I}$, $\alpha_{i}^{II}$, and $\alpha_{f}$, see (\ref{eq:F1_alpha_iI}) and (\ref{eq:F1_alpha_iII})) and, for direct comparison, the appropriately adjusted entrainment coefficients used in BK00, SH14, and \citet{debugne_hunt_2016} (who used same entrainment coefficients of SH14) are also plotted. a) The entrainment coefficients of the upflow at the inner boundary ($\alpha_{i}^{I}$ and $\alpha_{i}^{II}$, shown as blue dashed line and solid line, respectively), the entrainment coefficients of BK00 presented in adjusted form (lines with mark, using velocity scales from our DNS and taking the entrainment constants from BK00, namely $\alpha_{i_{00}} = 0.085$ and $\alpha_{f_{00}} = 0.147$, see (\ref{eq:F1_EI}) and (\ref{eq:F1_EII})), and the value used in both SH14 and \citet{debugne_hunt_2016}, $\alpha_{i}=0.06$ (vertical black line). The vertical dashed line marks zero coefficient. b) The entrainment coefficient of the downflow at the outer (orange line) calculated from (\ref{eq:F1_alpha_iI}) and (\ref{eq:F1_alpha_iII}) --- which give the identical result, and the entrainment coefficients used in BK00 and SH14 (thin black solid line and dashed line, respectively).}
     \label{fig:entrainmentcoeff}
 \end{figure}

Examination of figure~\ref{fig:entrainmentcoeff}, which plots our measurements of the entrainment coefficients and those used in fountain models thus far, challenges the validity of the entrainment assumptions that underpin the important contributions to fountain modelling of BK00, SH14, and \citet{debugne_hunt_2016}. The figure plots the values of $\alpha_{i}$ and $\alpha_{f}$, according to both entrainment formulations, i.e. calculated from our DNS data according to (\ref{eq:F1_alpha_iI}) and (\ref{eq:F1_alpha_iII}). Firstly, comparison of our data to that of \citet{Milton2022} yields similar values --- \citet{Milton2022}, to our knowledge, being the only other study to report the vertical variation of entrainment coefficient at the inner boundary of highly forced fountains ($Fr \geq 15$) based on measurements (therein under entrainment formulation EI and reported as a variation with the local Richardson number of the upflow). Both studies determine that the entrainment coefficient changes sign at a particular height within the upflow. Secondly, irrespective of the entrainment formulation chosen, the notable variation in our measurements of $\alpha_{i}$ and $\alpha_{f}$ with height renders the assumption of constant $\alpha_{f}$ in BK00, and constant $\alpha_{i}$ and $\alpha_{f}$ in SH14, as questionable. Thirdly, (\ref{eq:F1_EI}) and (\ref{eq:F1_EII}) provide the relationships that must hold true for the entrainment modelling at the inner boundary of BK00 to be valid. 

Figure~\ref{fig:entrainmentcoeff}(a) plots the LHS of \eqref{eq:F1_EI} and \eqref{eq:F1_EII}, i.e. the equivalent entrainment coefficients calculated from our data, namely, $\alpha_i^I$ and $\alpha_i^{II}$. In addition, it shows the profiles that follow from the existing theory, by substituting the constants used in BK00, i.e. $\alpha_{i_{00}} = 0.085$ and $\alpha_{f_{00}} = 0.147$ into the RHS of (\ref{eq:F1_EI}) and (\ref{eq:F1_EII}), and using our measured values of the velocity scales.
It is clear from the figure that these relationships do not hold true for the values taken; furthermore, we have tested various other choices for the entrainment constants within the BK00 formulation and there is an equally poor agreement for all physically reasonable entrainment constants. These facts highlight that an integral model of fountains, that captures the exchanges within the fountain, is yet to be developed, or at least parameterised appropriately. Integral models, to date, have been parameterised based on matching a single measurable statistic of fountain behaviour, the fountain height \citep[albeit sometimes both the initial and steady rise heights, e.g.][]{debugne_hunt_2016}, but our results suggest that no integral model has even been able to qualitatively captured the appropriate exchanges within fountains. As such, the agreement with the calibrated metric might be unique and all other fountain metrics from such models may be questionable. We have shown, \S\ref{sec:F1_ent}, that the total buoyancy exchanges, and the mean volume entrainment at the inner boundary, broadly follow a similar dependence of the vertical coordinate (see figures~\ref{fig:F1_entrainment}(a) and (g)); combined with figure~\ref{fig:entrainmentcoeff}, it shows that fountain models are yet to provide the capability to predict these fundamental exchanges which must affect the total weight of fountain fluid supported by a given source momentum flux. This highlights a major research challenge in fountain modelling, future models of fountains should seek validation via multiple statistics, not a single criterion (e.g. fountain height) as has been carried out previously.

In addition, by their very construction, bulk averaged fountain models (e.g. BK00 and SH14) did not account for the turbulent exchanges within fountains, e.g. turbulent momentum entrainment and turbulent buoyancy entrainment. We have shown in \S \ref{Sec:F1_Integral_quantities} and \S \ref{sec:F1_ent} that there is significant integral vertical turbulent momentum flux ($M'$, see figure~\ref{fig:F1_fluxes}(c)), and the turbulent momentum exchanges and turbulent buoyancy exchanges at the inner boundary (turbulent component of $m_i$ and $f_i$, respectively, see figure~\ref{fig:F1_entrainment}(d) and (g)). Hence, we challenge future models of fountains to account for these exchanges driven by turbulence, at the very least implicitly, within their parameterisation. 

\begin{figure}
    \centering
    \includegraphics{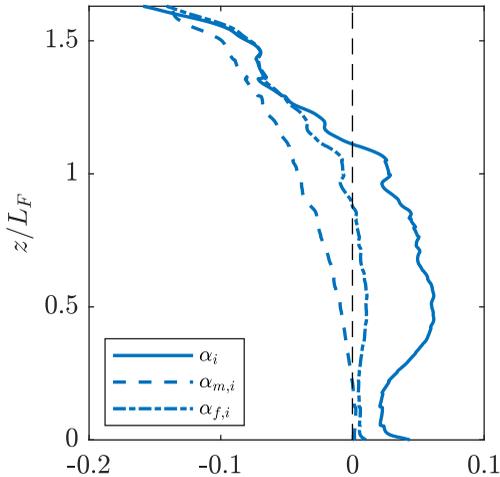}
    \caption{The entrainment coefficients for the upflow from our data: volume entrainment $\alpha_i$ from (\ref{eq:F1_EII}), momentum $\alpha_{m,i}$ obtained from the budget (\ref{eq:F1_mcoeff}) and buoyancy $\alpha_{f,i}$ from the budget (\ref{eq:F1_fcoeff}); i.e. all consistent with BK00's EII formulation.}
    \label{fig:T_alphas}
\end{figure}

Following the BK00 model, but considering entire budgets (we choose to consider only EII, a choice of little consequence), the total momentum entrainment (\ref{eq:F1_dMu}) and total buoyancy entrainment (\ref{eq:F1_dFu}) are
\begin{subequations}
\begin{align}
\label{eq:F1_mcoeff}
m_i &= 2 \alpha_{i,m} r_u w_u^2 \, ,\\    
\label{eq:F1_fcoeff}
f_i &= 2 \alpha_{i,f} r_u w_u b_u \, .
\end{align}
\label{eq:turbulentcoefficient}
\end{subequations}

Figure~\ref{fig:T_alphas} shows entrainment coefficients for momentum $\alpha_{i,m}$ and buoyancy $\alpha_{i,f}$ given by (\ref{eq:turbulentcoefficient}), respectively. The figure, for contrast, also re-plots $\alpha_i^{II}$, from (\ref{eq:F1_EII}), the plot clearly highlights that taking the same entrainment coefficient to parameterise exchanges of volume, momentum and buoyancy is inappropriate. More research is clearly needed to determine a suitable route to developing appropriate models of turbulent fountains; our leading suggestion is that the latest understanding of transports across both TNTIs and TTIs be considered in depth. Note that we do not present the entrainment coefficients decomposed into mean and turbulent components due to the lack of self-similarity of the radial profiles within the fountain \citep[\textit{cf.}][]{Milton2022}.

\section{Conclusion} \label{Sec: F1_conclusion}
The structure of a forced fountain was investigated by decomposing the fountain into regions based on two distinct internal boundaries. The first comprised the classical decomposition into an upflow and a downflow, via the internal boundary $r_i(z)$ characterised by the loci of zero vertical velocity. The second exploited a separatrix, a streamline $r_s(z)$ that forms the boundary between the flow emitted from the source and entrained from the environment. Both of these boundaries are turbulent/turbulent interfaces (TTIs), which can have both mean and turbulent contributions to entrainment across their boundary. For the internal boundaries investigated here, we showed that for the boundary $r_i$, entrainment fluxes are large and consist both of mean and turbulent components. However, the boundary $r_s$ by construction exhibits zero mean entrainment, since it is a streamline; hence, only turbulent exchanges can occur across.

The findings reported herein indicate that turbulent entrainment fluxes across the internal boundary are of similar magnitude to, and sometimes larger than, the mean entrainment fluxes. This has important implications for the modelling of turbulent entrainment. Indeed, the classical view of entrainment considers mean entrainment only, which then implies that an associated entrainment velocity can be defined across the interface, and entrainment of all other quantities (e.g. momentum, buoyancy) can then simply be obtained by multiplying that entrainment velocity by some local characteristic value of that quantity. The DNS data broadly supported this observation, in that the vertical variation of the fluxes of mean volume entrainment and mean buoyancy entrainment across $r_i$ exhibited a similar form (note the mean momentum entrainment which was zero by construction on $r_i$). However, the vertical variation of turbulent entrainment fluxes across $r_i$ exhibits an entirely different form to the mean entrainment fluxes and therefore requires different modelling. Even for the `classical' (mean volume) entrainment flux, the agreement between our DNS data and existing models was poor when the entrainment flux was normalised by suitable local characteristic scales, suggesting that parameterisation following MTT and deploying a constant entrainment coefficient is inappropriate for fountains and needs to be reconsidered. We further showed that different magnitudes of values are required for the respective entrainment coefficients to appropriately parameterise either mean entrainment, turbulent momentum entrainment, or turbulent buoyancy entrainment. This reinforces our suggestion that alternative modelling strategies, perhaps guided by more detailed studies of the transport across TNTIs and TTIs, might be fruitful if predictions of fountain behaviour, beyond their rise heights, are required. This is of relevance to many applications of fountains, for example, those produced within Under Floor Air Distribution Systems \citep{Linden2005} where predictions of fountain rise height only determine the vertical extent within the room that the cool air might rise; the cooling effect that occupants feel is further affected by the temperature of the rising and falling cooler air --- this depends both on the fountain's mixing with the environment and the mixing within the fountain's internal structure.

Finally, we assert that without greater regard to the modelling of the fountain cap region, insufficient progress might be made. In this region the flow is entirely dominated by the formation and collapse of large-scale turbulent structures --- characterised by periodic fluctuations in fountain rise height and responsible for the dark bands typically seen in time series images of fountain \citep[e.g. see figure~\ref{fig:F1_instantaneous_integral_buoyancy} and][]{burridge_hunt_2012}. Fountain modelling to date has focused on models that reflect only quantities of the mean flow --- our results suggest that both turbulent exchanges, and the transient nature of the fountain cap region, may be of importance in future modelling efforts. 

\section*{Acknowledgements}
The computations were performed on the UK National Supercomputing Service ARCHER2 and were made possible by the EPSRC-funded UK Turbulence Consortium (grant reference EP/R029326/1).

\section*{Declaration of interests}
The authors report no conflict of interest.

\appendix
\section{Comparison with smaller flow domain} \label{App: F1_dom_size}

To validate that the effect of the domain size, in particular the side-wall periodic boundary condition, was insignificant in our results we simulated identical source conditions in a `small' domain (of volume $(100r_0)^3$ and grid size $1024^3$); we refer to the domain used to simulate all other results presented herein as `standard'. 

\begin{figure}
	\centering{\includegraphics{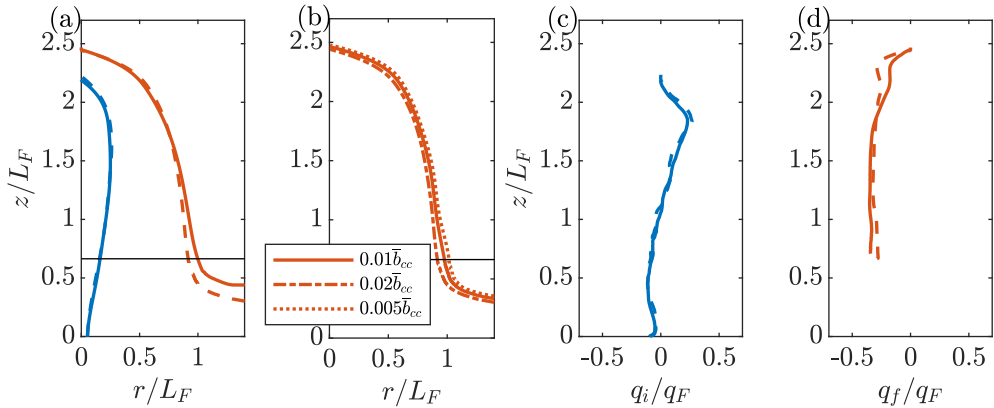}}	
	\caption{(a) Fountain boundaries, inner boundary $r_i$ (blue) and outer $r_f$ (orange) both from the standard domain (solid lines) and a small domain (dashed lines). (b) From the standard domain, the fountain outer boundary using a buoyancy threshold of $ 0.01\,\overline{b}_{cc}$ (solid line), $ 0.02\,\overline{b}_{cc}$ (dot-dashed line), and $ 0.005\,\overline{b}_{cc}$ (dotted line). The horizontal black line marks $z/L_F=0.66$ below where the data of downflow is discarded. Mean volume entrainment at inner boundary $q_i$ (c) and at outer boundary $q_f$ (d) of the standard domain (solid lines) and small domain (dashed lines) respectively.}
	\label{fig:boundaryconverge}
\end{figure}

Figure~\ref{fig:boundaryconverge}(a) plots the fountain inner boundary, $r_i$, and outer boundary, $r_f$, for simulations in both the standard and small domains. The effects of domain size on the fountain boundaries are negligible, particularly when one notes that we discard all results for the downflow (including $r_f$) below the height $z/L_F = 0.66$, with the typical change in the radial location of the boundaries with domain size being less than 7\%. Moreover, by integrating the volume of fountain $V = \int_0^{z_f} 2 \pi r_f(z) dz= 82\,730\,r_0^3 \approx 8.70 L_F^3$, the fountain occupies around $3\%$ of the standard domain, suggesting the standard domain is of an appropriate size.
 
To check the location of the outer boundary is relatively insensitive to the choice of buoyancy threshold ($\overline{b} = 0.01 \overline{b}_{cc}$), figure~\ref{fig:boundaryconverge}(b) plots the outer boundary using both a larger threshold, $\overline{b} = 0.02 \overline{b}_{cc}$, and a smaller threshold, $\overline{b} = 0.005 \overline{b}_{cc}$. The figure evidences that the outer boundary is indeed insensitive on the choice threshold, within our range of interest and especially above $z/L_F= 0.66$. We therefore conclude that our outer boundary, $r_f$, contains the vast majority of fountain fluid and is appropriate. 

To ensure the entrainment at the boundaries is also less influenced by the domain, figure \ref{fig:boundaryconverge}(c) and \ref{fig:boundaryconverge}(d) plot the mean volume entrainment $q_i$ and $q_f$ of two domains respectively. The lines agree well though there is a relatively greater difference in $q_f$ near the fountain top. By calculating the integral volume entrainment of the fountain cap $Q_{cap}$, the value of the small domain is about $-0.19Q_F$, not far from the value of the standard domain which is $-0.22Q_F$.

\begin{figure}
    \centering
    \includegraphics[scale=0.35]{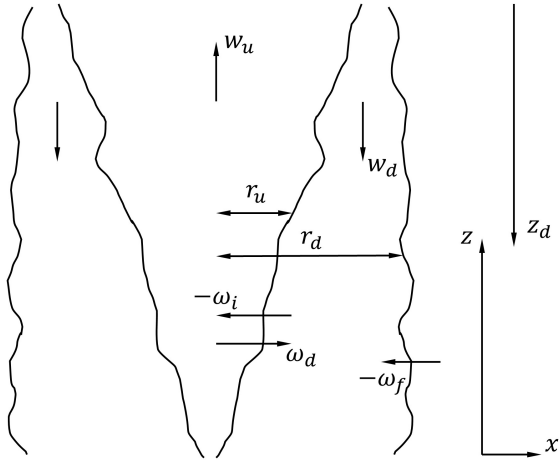}
    \caption{Fountain model BK00, with the arrows indicating the direction of variables.}
    \label{fig:BKmodel}
\end{figure}

\section{The integral equations of BK00} \label{App: F1_BKeq}
In order to provide a direct comparison of the entrainment coefficients of the well-established BK00 model to those presented herein, i.e. those apparent on rigorous application of Reynolds averaging of the governing equations, we first present the BK00 model in our notation (the notation introduced in \S \ref{Sec:F1_equations}) within figure~\ref{fig:BKmodel}. The velocities $\omega_{i}$, $\omega_{d}$ characterise the two-way entrainment across the inner boundary, and $\omega_{f}$ characterises the entrainment at the outer boundary. As these velocities are defined under our outward horizontal direction, $\omega_{i}$ and $\omega_{f}$ here should associate with a negative sign to represent the entrainment velocity into the respective flow.

In the model of BK00, the integral fluxes are defined equivalently to those defined in (\ref{eq.F1_charu})--(\ref{eq.F1_chard}) and note that the BK00 model assumed (top-hat) uniform fluid properties and velocities in each flow region.  BK00 directly applied integral conservation laws for volume, momentum and buoyancy fluxes which resulted in
\begin{subequations}
\label{eq:F1_BKeqs}
\begin{align}
\label{eq:F1_BKQu}
    \frac{\d Q_u}{\d z}   &= 2r_u (-\omega_{i}) - 2r_u \omega_{d} \, ,\\
    \frac{\d Q_d}{\d z_d} &= 2r_d (-\omega_{f}) - 2r_u (-\omega_{i}) + 2r_u \omega_{d} \, ,\\
\label{eq:F1_BKMu}
    \frac{\d M_u}{\d z}   &= B_u + 2r_u (-\omega_{i}) (-w_d) - 2r_u \omega_{d} w_u \, ,\\
\label{eq:F1_BKMd}
    \frac{\d M_d}{\d z_d} &= B_d + 2r_u \omega_{d} (-w_u) - 2r_u (-\omega_{i}) w_d \, ,\\
    \frac{\d F_u}{\d z}   &= 2r_u (-\omega_{i}) (-b_d) - 2r_u \omega_{d} b_u \, ,\\
    \frac{\d F_d}{\d z_d} &= 2r_u \omega_{d} (-b_u) - 2r_u (-\omega_{i}) b_d  \, .
\end{align}
\end{subequations}

As is typical, BK00 considered mean integral fluxes only, and thus the terms $M_u', P_u, F_u',M_d', P_d$ and $F_d'$ in (\ref{eq:F1_du}) and (\ref{eq:F1_dd}) are identically zero. Momentum and buoyancy entrainment from the environment is assumed to be negligible. As discussed in \S \ref{sec:F1_ent}, BK00 use two-way entrainment at the inner boundary, which then implicitly contains the Leibniz contributions to the exchange.

We note that the BK00 model assumes top-hat conditions which imply that the upflow region has a width $r_u = Q_u/M_u^{1/2}$; in reality, the profiles are not top-hat and thus the actual internal boundary is located at a greater radial position and equivalent to $r_i \approx 1.25 r_u$. 
Another consequence of the top-hat assumption is that the momentum and buoyancy on the boundary are equivalent to their values in the upflow or downflow.

\bibliographystyle{jfmF}
\bibliography{References}


\end{document}